\documentclass[aps,prd,superscriptaddress,notitlepage,nofootinbib,10pt]{revtex4-1}

\usepackage{graphicx,amsfonts,amsmath,amssymb,amstext}
\usepackage{float,wrapfig}
\usepackage{subfigure, psfrag}
\usepackage{dsfont}
\usepackage{color}
\usepackage{bm}

\definecolor{purple}{rgb}{0.8,0,0.6}
\definecolor{darkgreen}{rgb}{0.00,0.6,0.00}

\makeatletter

\newcommand*{\Rmnum}[1]{\text{\expandafter\@slowromancap\romannumeral #1@}}
\makeatother

\begin{document}

\title{Hydrodynamic modes in a magnetized chiral plasma with vorticity}

\author{D.~O.~Rybalka}
\affiliation{Department of Physics, Arizona State University, Tempe, Arizona 85287, USA}

\author{E.~V.~Gorbar}
\affiliation{Department of Physics, Taras Shevchenko National Kiev University, Kiev, 03680, Ukraine}
\affiliation{Bogolyubov Institute for Theoretical Physics, Kiev, 03680, Ukraine}

\author{I.~A.~Shovkovy}
\affiliation{Department of Physics, Arizona State University, Tempe, Arizona 85287, USA}
\affiliation{College of Integrative Sciences and Arts, Arizona State University, Mesa, Arizona 85212, USA}

\date{January 24, 2019}

\begin{abstract}
By making use of a covariant formulation of the chiral kinetic theory in the relaxation-time approximation, 
we derive the first-order dissipative hydrodynamics equations for a charged chiral plasma with background
electromagnetic fields. We identify the global equilibrium state for a rotating chiral plasma confined to a 
cylindrical region with realistic boundary conditions. Then, by using linearized hydrodynamic equations, 
supplemented by the Maxwell equations, we study hydrodynamic modes of magnetized rotating chiral 
plasma in the regimes of high temperature and high density. We find that, in both regimes, dynamical 
electromagnetism has profound effects on the spectrum of propagating modes. In particular, there are 
only the sound and Alfv\'en waves in the regime of high temperature, and the plasmons and helicons at high 
density. We also show that the chiral magnetic wave is universally overdamped because 
of high electrical conductivity in plasma that causes an efficient screening of charge fluctuations. The 
physics implications of the main results are briefly discussed. 
\end{abstract}

\maketitle

\section{Introduction}
\label{sec:intro}

Chiral plasmas can be realized in a number of physical systems, ranging from degenerate forms of 
dense matter in compacts stars \cite{Weber:2004kj,Yamamoto:2015gzz} to hot plasmas in the early universe 
\cite{Joyce:1997uy,Vallee:2011zz,Durrer:2013pga} and heavy-ion collisions \cite{Liao:2014ava,Kharzeev:2015znc,Huang:2015oca}. 
A nonelectromagnetic single-chirality plasma can be formed by neutrinos trapped in a dense nuclear 
medium formed during the early stages of the supernova explosions \cite{Burrows:1986me}. The ultrarelativistic matter 
inside the jets of accreting black holes is yet another example of a chiral plasma in astrophysics \cite{McKinney:2006tf}. 
In condensed matter physics, pseudorelativistic analogues of chiral plasmas are realized by the 
low-energy electron quasiparticles in Dirac and Weyl materials \cite{Vafek:2013mpa,Burkov:2015hba}. 

As numerous theoretical studies suggest, the corresponding chiral forms of matter could exhibit unusual 
phenomena that have roots in quantum anomalies. A partial list of such phenomena includes the chiral 
magnetic \cite{Fukushima:2008xe}, chiral separation \cite{Vilenkin:1980fu,Metlitski:2005pr}, and chiral 
vortical \cite{Vilenkin:1978hb,Vilenkin:1979ui,Vilenkin:1980zv,Erdmenger:2008rm,Banerjee:2008th,
Son:2009tf,Sadofyev:2010pr,Neiman:2010zi,Landsteiner:2011cp} effects. Such effects not only 
modify the transport properties of relativistic matter but also give rise to new types of low-energy 
collective modes \cite{Kharzeev:2010gd,Jiang:2015cva,Chernodub:2015gxa,Frenklakh:2015fzc}. 
The latter, in turn, could lead to a range of anomalous observables 
in heavy-ion collisions; see Refs.~\cite{Liao:2014ava,Kharzeev:2015znc,Huang:2015oca} 
for reviews. (For similar studies in the context of Dirac/Weyl materials, see Refs.~\cite{Pellegrino:2015qtu,
Gorbar:2016sey,Gorbar:2016vvg,Long:2017xnj}.) It is remarkable that the anomalous features 
can be described by using semiclassical approaches such as kinetic theory and hydrodynamics. 
It is not surprising, therefore, that the corresponding chiral formulations of the kinetic theory 
\cite{Son:2012wh,Stephanov:2012ki,Son:2012zy,Manuel:2014dza} and hydrodynamics 
\cite{Son:2009tf,Sadofyev:2010pr,Neiman:2010zi} attracted a lot of attention recently. 

While the ideal chiral hydrodynamics can be obtained from general principles and symmetries alone
\cite{Son:2009tf,Sadofyev:2010pr,Neiman:2010zi}, the inclusion of dissipative effects poses a much 
harder problem. Recently, this problem was partially resolved by us in Ref.~\cite{Gorbar:2017toh},
where the second-order dissipative hydrodynamics for a chiral plasma made of neutral particles
was formulated. Starting from the chiral kinetic theory (CKT) in the relaxation-time approximation, we 
derived the corresponding set of hydrodynamic equations by using the Chapman-Enskog method 
and a special type of self-consistent truncation \cite{Denicol:2010xn,Jaiswal:2013npa,Jaiswal:2013vta,
Jaiswal:2015mxa}. 

In connection to the CKT, which served as a starting point in the derivation of dissipative hydrodynamics 
in Ref.~\cite{Gorbar:2017toh}, we should note that its original formulation 
\cite{Son:2012wh,Stephanov:2012ki,Son:2012zy} did not have an explicit Lorentz covariance and 
did not account for the effects of collisions. Both of these facts presented serious limitations for practical 
applications of the CKT. The issue of Lorentz covariance was partially addressed in the derivations 
based on the Wigner-function approach in Refs.~\cite{Gao:2012ix,Chen:2012ca}. A nontrivial realization 
of the Lorentz invariance was further illuminated in Ref.~\cite{Chen:2014cla}, where the need for 
the so-called ``side jumps" in scattering processes was advocated. The structure of collision 
terms was further investigated in Ref.~\cite{Chen:2015gta}, but its generalization to the case with 
background electromagnetic fields remained unclear. A truly covariant formulation of the CKT 
in background electromagnetic fields that treats collisions self-consistently was obtained recently 
in Refs.~\cite{Hidaka:2016yjf,Hidaka:2017auj} (see also Ref.~\cite{Huang:2018wdl}).

In this study, by making use of the covariant formulation of the CKT \cite{Hidaka:2016yjf,Hidaka:2017auj}
in the relaxation-time approximation, we derive the equations of dissipative hydrodynamics for a magnetized 
chiral plasma with nonzero vorticity and study its collective modes. It might be appropriate to note 
that the use of the hydrodynamic description is the relevant and most appropriate in the long-wavelength regime of a chiral 
plasma. Unlike the chiral kinetic theory, for example, it naturally incorporates the oscillations of fluid flow. 
Previously, some hydrodynamic modes in the presence of background magnetic fields and vorticity have been discussed 
in Refs.~\cite{Yamamoto:2015ria,Abbasi:2015saa,Kalaydzhyan:2016dyr,Abbasi:2016rds,Hattori:2017usa},
but not always with sufficient rigor. (For studies of hydrodynamic modes in Dirac/Weyl materials, see also 
Ref.~\cite{Gorbar:2018nmg}.) One of the common limitations is the use of the approximation with 
nondynamical background fields. While many new exotic modes are predicted, it is unknown whether 
any of them would survive when dynamical electromagnetism is properly accounted for. The novel and 
distinctive feature of the present study is a self-consistent treatment of electromagnetic effects in a 
charged chiral plasma. As we demonstrate, dynamical electromagnetism plays an essential 
role in shaping the physical properties of collective modes (with and without rotation in plasma). 
One of its most dramatic consequences is the absence of the chiral magnetic wave, which becomes
strongly overdamped due to high conductivity of hot and/or dense plasmas.

This paper is organized as follows. In Sec.~\ref{sec:theoretical_framework}, we start by deriving the 
complete set of dissipative chiral hydrodynamics equations from the CKT. The global equilibrium 
state of a uniformly rotating chiral plasma in a background magnetic field is discussed in Sec.~\ref{sec:Equilibrium}. 
The linearized hydrodynamic equations for deviations of hydrodynamic variables from their
equilibrium values are derived in Sec.~\ref{sec:Linearized}. The studies of hydrodynamic modes in the 
high-temperature and high-density regimes of a chiral magnetized rotating plasma are presented in 
Secs.~\ref{sec:Modes-highT} and \ref{sec:Modes-highMu}, respectively. In both regimes, we 
investigate the role of dynamical electromagnetism on the properties of the modes. The general 
discussion and the summary of our main results are given in Sec.~\ref{sec:Summary}. Several 
appendixes at the end of the paper contain some technical details and useful reference material. 

Throughout this paper we use the units with $c=1$, but keep track of the Planck constant $\hbar$ 
explicitly. Also, the convention for the Minkowski metric is $g_{\mu\nu}=\mbox{diag}(1,-1,-1,-1)$
and the Levi-Civit\`a tensor $\epsilon^{\mu\nu\alpha\beta}$ is defined so that $\epsilon^{0123} = 1$.

\section{Theoretical framework}
\label{sec:theoretical_framework}

In this section we derive a closed system of chiral hydrodynamic equations from the covariant 
version of the CKT \cite{Hidaka:2016yjf,Hidaka:2017auj}. The latter was obtained from the 
quantum-field theoretic formulation of massless QED by applying the Schwinger-Keldysh formalism. 
In the corresponding description, the lesser/greater propagators are directly connected to the 
Wigner function. Unlike the early heuristic approaches based on the Wigner function for noninteracting 
fermions \cite{Gao:2012ix,Chen:2012ca}, the derivation in Refs.~\cite{Hidaka:2016yjf,Hidaka:2017auj} 
not only accounts for background electromagnetic fields but also includes the effects of interactions. 
A similar description might also be possible by using the on-shell effective field theory that was recently 
proposed in Ref.~\cite{Carignano:2018gqt}. 

When dealing with charged plasmas in the hydrodynamic regime, the electromagnetic fields should 
be treated as fully dynamical, even in the static and steady-state cases. This implies that the 
commonly used background field approximation is not reliable in investigations of hydrodynamic 
modes. Therefore, in our study below, we supplement the equations for the hydrodynamic variables
with the coupled Maxwell equations for the 
electromagnetic fields. As we will demonstrate below, such a self-consistent treatment will be 
important not only for the correct description of the hydrodynamic modes but also for identifying
the global equilibrium state in a magnetized relativistic plasma with nonzero vorticity. 

In order to capture dissipative effects, we should start our derivation from the CKT that takes particle 
interactions into account. Instead of introducing a complete particle collision integral, however, 
we will utilize the so-called relaxation-time approximation. From the viewpoint of the resulting hydrodynamic 
description, which assumes a local thermal equilibrium on sufficiently short distance scales, this should 
be an adequate approximation. It should be noted, however, that enforcing Lorentz invariance 
in the relaxation-time approximation is far from trivial \cite{Hidaka:2018ekt}. Traces of this problem 
will also show up in our derivation below where we will find that the consistency of hydrodynamic equations 
(\ref{eq:cont-j})--(\ref{eq:cont-T}) requires a special choice of the reference frame.

In order to set up the notations, let us start from a short introduction into the formalism 
used in Refs.~\cite{Hidaka:2016yjf,Hidaka:2017auj}. By definition, the Wigner function 
of Weyl fermions is given by $W_{\alpha\beta}(p,x) = (2\pi)^3\langle \psi^\dagger_\beta(x)
\delta^4(\hat p - p) \psi_\alpha(x) \rangle$, where $\alpha,\beta=1,2$ are the spinor indices, 
$\hat p_\mu \equiv \frac{i}{2}\left(\partial_\mu-\partial_\mu^\dagger\right)$, and $\psi(x)$ is 
a second quantized Weyl spinor of a given chirality. (For a general overview of the Wigner 
function formalism, see Ref.~\cite{Vasak:1987um}.) Since the Wigner function is a matrix 
in the spinor space, it can be conveniently represented in terms of  the Pauli matrices, i.e., 
$W(p,x) = W^\mu(p,x)\sigma_\mu$, where $\sigma^\mu=(1,\vec{\sigma})$. 
The four-vector $W^\mu(p,x)$ is related to the phase-space density of the number density 
current of chiral fermions with momentum $p$ at position $x$. 
As we will see below, therefore, one can also relate the trace of the Wigner function to 
the (quasi)classical distribution function of chiral particles. 

In general, we will consider chiral plasmas that are made of fermions of both chiralities, 
denoted by $\lambda = \pm 1$, where the plus (minus) sign corresponds to the right-handed 
(left-handed) fermions. In the following, we will assume that fermions of each chirality 
are described by independent Wigner functions or, in other words, that $W(p,x)$ depends on 
the chirality index $\lambda$. For simplicity of the notation, however, such a dependence will 
not be displayed explicitly, although it will always tacitly be assumed.

The quasiclassical solution for the Wigner function can be obtained iteratively by using the 
expansion in powers of $\hbar$ \cite{Hidaka:2016yjf,Hidaka:2017auj}. The zeroth order result, 
in particular, is given by $W_{(0)}^\mu(p,x) = p^\mu\delta(p^2) f(p,x)$, where function $f(p,x)$ 
satisfies the relativistic Boltzmann equation for an ideal gas in a background electromagnetic 
field, i.e., $p^\mu\mathcal{D}_\mu f(p,x) = 0$. Here, the phase space derivative is defined by 
$\mathcal{D}^\mu = \partial/\partial x^\mu - eF^{\mu\nu} \partial/\partial p^\nu$ and $F^{\mu\nu}$ 
is the electromagnetic field strength tensor. The function $f(p,x)$ can be interpreted as
a particle distribution function.

To the linear order in $\hbar$, the distribution function $f(p,x)$ satisfies the following covariant  
CKT equation \cite{Hidaka:2016yjf,Hidaka:2017auj}:
\begin{equation}
\label{eq:CKT-equation}
	\mathcal{D}_\mu W^\mu(p,x) 
	= \delta(p^2) p\cdot C 
	+ \lambda\hbar e\tilde{F}^{\mu\nu}C_\mu p_\nu \delta'(p^2),
\end{equation}
where $\tilde F^{\mu\nu} = \frac{1}{2}\epsilon^{\mu\nu\alpha\beta}F_{\alpha\beta}$ is the dual of the 
field strength tensor and $W^\mu(p,x)$ is the Wigner function with the corrections up to subleading 
order. The explicit form of the latter is given in terms of $f(p,x)$ as follows:
\begin{equation}
\label{eq:wigner-first-order}
	W^\mu(p,x) \equiv p^\mu \delta(p^2)f + \lambda\hbar S^{\mu\nu} \delta(p^2) (D_\nu f - C_\nu) 
	+ \lambda\hbar e\tilde{F}^{\mu\nu} p_\nu \delta'(p^2) f + O(\hbar^2)
\end{equation}
Here, $S^{\mu\nu} = \frac{1}{2} \epsilon^{\mu\nu\alpha\beta} p_\alpha u_\beta/(p\cdot u)$ 
is a particle spin tensor and $C^\mu$ is a collision operator which will be defined later. Note that 
the spin tensor is expressed in terms of a timelike four-vector $u^\mu(x)$ that defines the local frame.

It is instructive to review the physical meaning of the individual terms on the right-hand side 
of Eq.~(\ref{eq:wigner-first-order}). The first one, which gives the zeroth order result, 
describes the classical free particle streaming. The second term captures the spin-orbit interaction 
and the effects of collisions. It is critical for the chiral vortical effect and the current connected 
with side jumps (see also Ref.~\cite{Chen:2014cla}). The third term on the right-hand side of 
Eq.~(\ref{eq:wigner-first-order}) describes the interaction of the magnetic moment with the 
background field and is responsible for the chiral magnetic effect. We also note that the first 
two terms enforce the conventional massless dispersion relation for chiral fermions, i.e., 
$p^2=0$, whereas the last one accounts for quantum corrections to the dispersion relation. 

As mentioned earlier, we will treat the collisions in the CKT by employing the relaxation-time 
approximation. In this approximation, the Lorentz covariant collision operator has a particularly simple form
$C^\mu = - u^\mu (f-f_\textrm{eq})/\tau$, where $\tau$ is the relaxation time and $f_\textrm{eq}(p,x)$ is the 
equilibrium distribution function \cite{Hidaka:2018ekt}. In this case, $S^{\mu\nu} C_\nu \equiv 0$ and, 
therefore, the side-jump term in the Wigner function vanishes. After taking into account Eq.~(\ref{eq:wigner-first-order}), 
the CKT equation (\ref{eq:CKT-equation}) takes a rather simple  form, i.e., 
\begin{equation}
\label{eq:CKT-equation-2}
	\mathcal{D}_\mu W^{\mu} = - \frac{u_\mu (W^\mu - W^\mu_\textrm{eq})}{\tau},
\end{equation}
where $W^\mu_\textrm{eq}$ is the Wigner function, associated with the equilibrium state. For the equilibrium 
distribution function in the local frame of the fluid, we will assume the usual Fermi-Dirac distribution, i.e.,  
\begin{equation}
\label{eq:f_equilibrium}
	f_\textrm{eq}(p,x) = \frac{1}{1 + e^{\textrm{sign}(p_0)(\varepsilon_p-\mu_\lambda)/T}}.
\end{equation}
Note that the chirality index $\lambda$ (not to be confused with a Lorentz index) is shown 
explicitly in the chemical potential $\mu_\lambda \equiv \mu+\lambda \mu_5$, where $\mu$ is the 
fermion number chemical potential and $\mu_5$ is the chiral chemical potential.
By definition, the particle dispersion relation is given by 
\begin{equation}
\label{eq:p-dispersion}
\varepsilon_p = u_\mu p^\mu + \frac{\lambda\hbar}{2} \frac{p\cdot\omega}{p\cdot u}, 
\end{equation}
where $\omega^\mu = \frac{1}{2}\epsilon^{\mu\nu\alpha\beta}u_\nu\partial_\alpha u_\beta$ is the local 
vorticity \cite{Hidaka:2016yjf,Hidaka:2017auj}. The vortical term in Eq.~(\ref{eq:p-dispersion}) accounts 
for the spin-orbit coupling and, as suggested by its dependence on $\hbar$, has a quantum origin. 
In order to describe both particles (with $p_0>0$) and antiparticles (with $p_0<0$), we introduced the 
energy sign factor, $\textrm{sign}(p_0)$, in the exponent of the distribution function (\ref{eq:f_equilibrium}). 

As is clear from Eqs.~(\ref{eq:f_equilibrium}) and (\ref{eq:p-dispersion}), the local equilibrium state is determined by 
six independent parameters, i.e., the local temperature $T(x)$, the two local chemical potentials $\mu_\lambda(x)$ 
(for the two species of fermions with opposite chiralities), and three (out of total four)  independent
components of the local rest-frame velocity $u^\mu(x)$, constrained by the normalization condition 
$u^\mu u_\mu = 1$. These parameters fully describe the local thermodynamic state of a chiral plasma. 
It might be important to emphasize that we will treat the plasma, which is made of two types of particles 
of opposite chirality, as a one-component fluid. This means that the local equilibrium state is characterized 
by the same common temperature and fluid velocity independent of the chirality. However, we will allow for
the chemical potentials of opposite chirality fermions to be different. The corresponding hydrodynamic 
regime can be justified when the chirality-changing (anomalous) processes are sufficiently rare, i.e., 
when the rate of the thermal (kinetic) equilibration is much faster than the rate of the anomalous processes. 
From a theoretical viewpoint, this is a particularly interesting regime as it allows for chiral effects to be 
realized and observed in quasiclassical systems.  

In essence, the hydrodynamic equations are nothing else but the continuity equations for the conserved 
quantities. In the case of a charged chiral plasma, they are the fermion number and chiral 
charge currents, as well as the energy-momentum tensor. (Note that here we use the fermion 
number current $j^\mu$ instead of the electric current $j_{\rm el}^\mu \equiv e j^\mu$.) In terms of the 
chiral Wigner functions, the corresponding quantities are given as follows \cite{Vasak:1987um}:
\begin{eqnarray}
\label{eq:current-from-W}
	j^\mu &=& 2 \sum_\lambda \int\frac{d^4p}{(2\pi \hbar)^3} W^\mu,
	\\
\label{eq:current5-from-W}
	j_5^\mu &=& 2 \sum_\lambda \lambda \int\frac{d^4p}{(2\pi \hbar)^3} W^\mu,
	\\
\label{eq:Tmunu-from-W}
	T^{\mu\nu} &=& \sum_\lambda \int\frac{d^4p}{(2\pi \hbar)^3} (W^\mu p^\nu + p^\mu W^\nu).
\end{eqnarray}
It might be instructive to note that these expressions contain an additional 
dependence on the Planck constant $\hbar$ entering through the phase-space measure. 
Such a dependence is not connected with the use of the quasiclassical approximation for 
the Wigner function, but simply accounts for a correct number of microstates in any given 
macrostate. Interestingly, the same dependence on $\hbar$ in the phase-space measure 
should appear even in the classical theory, although it usually drops out from 
many classical observables and thermodynamic relations. In our study of a chiral plasma, 
however, an explicit dependence on $\hbar$ will remain in most thermodynamic quantities, 
including the particle and energy densities, the pressure, and transport coefficients. (For 
explicit expressions of some thermodynamic quantities, see Appendix~\ref{app:integrals}.)

Before proceeding further, it is useful to comment on several possible definitions of the local
rest frame $u^\mu(x)$ in chiral fluids. In relativistic hydrodynamics, the most common 
choices of local frames are the so-called Eckart frame, connected with the conserved 
charge (e.g., the fermion number or the electric charge) current ($u^\mu \parallel j^\mu$), 
and the Landau frame, connected with the energy flow ($u^\mu \parallel T^{\mu\nu}u_\nu$). 
In the case of chiral fluids, however, the preferred choice might be the so-called no-drag frame 
introduced in Refs.~\cite{Rajagopal:2015roa,Stephanov:2015roa,Sadofyev:2015tmb}. The local 
thermal equilibrium in the corresponding frame is described by the usual Fermi-Dirac 
distribution function. This is also the choice that we assume here. 

By making use of the local rest frame of the fluid $u^\mu$, it is convenient to decompose the 
currents in terms of the longitudinal and transverse components, i.e., 
\begin{eqnarray}
\label{eq:hydro_param}
	j^\mu &=& n u^\mu + \nu^\mu,
	\\
	j_5^\mu &=& n_5 u^\mu + \nu_5^\mu,
\end{eqnarray}
where $n=j^\mu u_\mu$ and $n_5 = j_5^\mu u_\mu$ are the fermion number and chiral charge densities, 
respectively. The transverse currents, $\nu^\mu = \Delta^{\mu\nu} j_\nu$ and $\nu_5^\mu = \Delta^{\mu\nu} j_{5,\nu}$, 
are obtained by removing the longitudinal components of the corresponding four-currents with the help 
of the projection operator $\Delta^{\mu\nu} \equiv g^{\mu\nu} - u^\mu u^\nu$. Here it might be appropriate 
to mention in passing that, in the case of chiral fluids, the currents $\nu^\mu$ and $\nu_5^\mu$ may  
contain not only the usual dissipative contributions but also anomalous nondissipative ones that originate
from quantum anomalies (see below). 

The analogous decomposition for the energy-momentum tensor reads
\begin{equation}
T^{\mu\nu} = \epsilon u^\mu u^\nu - \Delta^{\mu\nu} P + (h^\mu u^\nu + u^\mu h^\nu) + \pi^{\mu\nu},
\end{equation}
where $\epsilon = T^{\mu\nu} u_\mu u_\nu$ is the energy density, 
$P = \Delta_{\mu\nu} T^{\mu\nu}/3$ is the pressure, $h^\mu = \Delta^{\mu\alpha} T^{\alpha\beta} u_\beta$ 
is the energy flow (or, equivalently, the momentum density vector), and $\pi^{\mu\nu} = \Delta^{\mu\nu}_{\alpha\beta} T^{\alpha\beta}$ 
is the dissipative part of the energy-momentum tensor, which is defined in terms of the traceless four-index 
projection operator $\Delta^{\mu\nu}_{\alpha\beta} = \frac{1}{2}\Delta^\mu_\alpha\Delta^\nu_\beta + \frac{1}{2}\Delta^\mu_\beta\Delta^\nu_\alpha - \frac{1}{3}\Delta^{\mu\nu}\Delta_{\alpha\beta}$. 

We note that, by making use of the definition of the energy-momentum tensor (\ref{eq:Tmunu-from-W}) and 
the Wigner function (\ref{eq:wigner-first-order}), one can derive the well-known equation of state for an 
ideal gas of massless fermions: $P = \Delta_{\mu\nu} T^{\mu\nu}/3 = \epsilon/3$. In such an approximation, 
the speed of sound is $c_s=1/\sqrt{3}$. In a more realistic case of interacting massless fermions, the value of 
$c_s^2$ is expected to be somewhat smaller than $1/3$, but usually not much. Indeed, even in a strongly 
interacting quark-gluon plasma, the value of $c_s^2$ is found to be about $0.25$ to $0.3$ for almost all 
temperatures above the deconfinement transition \cite{Bazavov:2014pvz}. Moreover, $c_s^2$ increases 
with temperature and reaches within 10\% of the ideal gas value already at $T\simeq 350~\mbox{MeV}$.

In our study here, we address the qualitative features of collective modes and the role of dynamical 
electromagnetism in the chiral hydrodynamics framework. Since a specific value 
of the sound velocity in this context is not of much importance, it will be sufficient for us to use the 
simplest equation of state of an ideal massless gas. It will also be clear that none of our qualitative 
results will change if a more complicated equation of state is used.

By calculating the divergences of the current densities, defined in Eqs.~(\ref{eq:current-from-W}) and 
(\ref{eq:current5-from-W}) in terms of the Wigner function, and the energy-momentum 
tensor in Eq.~(\ref{eq:Tmunu-from-W}) and then taking also the CKT equation (\ref{eq:CKT-equation-2}) 
into account, we obtain the following relations:
\begin{eqnarray}
\label{eq:cont-j}	
\partial_\mu j^\mu &=& - \frac{1}{\tau} (n - n_\textrm{eq}),
	\\
\label{eq:cont-j5}	
\partial_\mu j_5^\mu +\frac{e^2}{8\pi^2\hbar^2} F^{\mu\nu} \tilde F_{\mu\nu} 
&=& - \frac{1}{\tau} (n_5 - n_{5,\textrm{eq}}),
	\\
\label{eq:cont-T}	
\partial_\nu T^{\mu\nu} - eF^{\mu\nu} j_\nu
	&=&  - \frac{u^\mu}{\tau} \left(\epsilon - \epsilon_\textrm{eq} + \frac{\hbar}{2} \omega_\nu (\nu_5^\nu - \nu^\nu_{5,\textrm{eq}}) \right) 
	- \frac{1}{\tau} \left( h^\mu - h^\mu_\textrm{eq} - \frac{\hbar}{4}\epsilon^{\mu\alpha\beta\gamma} u_\alpha \dot{u}_\beta (\nu_{5,\gamma} - \nu^\textrm{eq}_{5,\gamma}) \right),
\end{eqnarray}
where $\dot{u}_\beta = u^\alpha \partial_\alpha u_\beta$ and all equilibrium quantities on the right-hand 
sides are calculated by using the distribution function $f_\textrm{eq}(p,x)$ in Eq.~(\ref{eq:f_equilibrium}). 
Their explicit expressions are well known and are given in Appendix~\ref{app:integrals}. Of course, the 
correct forms of the corresponding continuity equations in the chiral plasma should have the vanishing 
right-hand sides. This is clearly not the case for Eqs.~(\ref{eq:cont-j})--(\ref{eq:cont-T}) derived from the 
CKT in the relaxation-time approximation. In fact, this is a well-known artifact of such an approximation 
\cite{Anderson-Witting}. The root of the problem lies in the equilibrium distribution function, which is used 
in the definition, but was not fully specified yet. The conservation laws are enforced by imposing the 
following self-consistency conditions \cite{Anderson-Witting}:
\begin{eqnarray}
\label{eq:matching-conditions-n}
	n &=& n_\textrm{eq},
	\\
\label{eq:matching-conditions-n5}
	n_5 &=& n_{5,\textrm{eq}}, 
	 \\
\label{eq:matching-conditions-e}
	\epsilon +\frac{\hbar}{2} \omega_\mu \nu_5^\mu &=& \epsilon_{\rm eq} + \frac{\hbar}{2} \omega_\mu \nu^\mu_{5,{\rm eq}}, 
	\\
\label{eq:matching-conditions-h}
	h^\mu -  \frac{\hbar}{4} \epsilon^{\mu\alpha\beta\gamma} u_\alpha \dot{u}_\beta \nu_{5,\gamma} &=& h^\mu_{\rm eq} - \frac{\hbar}{4} \epsilon^{\mu\alpha\beta\gamma} u_\alpha \dot{u}_\beta \nu_{5,{\rm eq},\gamma}.
\end{eqnarray}
These can be viewed as the defining relations for the local equilibrium parameters $T$, $\mu_\lambda$, and 
$u^\mu$ in terms of the local values of the fermion number density $n$, the chiral charge density $n_5$, the 
energy density $\epsilon$, and the momentum density $h^\mu$. Alternatively, the above conditions specify 
the local fermion number density, the chiral charge density, the energy density, and the momentum density, 
respectively, in terms of the local equilibrium values of $T$, $\mu_\lambda$, and $u^\mu$. 

Before proceeding further with the derivation, it is instructive to mention that the second term 
in Eq.~(\ref{eq:cont-j5}) describes the chiral anomaly, which explicitly breaks the conservation of the 
chiral charge. In principle, Eq.~(\ref{eq:cont-j5}) may also contain a similar anomalous contribution 
from the non-Abelian gauge fields. In fact, it is known that non-Abelian topological configurations could
play an important role in heavy-ion collisions. For example, they may produce metastable 
$\mathcal{P}$- and $\mathcal{CP}$-odd domains with a nonzero chiral imbalance that could be detected 
via the chiral magnetic effect \cite{Kharzeev:2004ey,Kharzeev:2007jp}. Additionally, the non-Abelian 
gauge configurations with parallel chromoelectric and chromomagnetic fields could play an important 
role in the early (``glasma") stage of heavy-ion collisions \cite{Lappi:2006fp}. In the hydrodynamic 
description used here, however, we do not include such effects explicitly. In the long-wavelength limit, 
they are captured effectively by inclusion of a nonzero chiral chemical potential.

The dissipative components of the currents and the energy-momentum tensor, i.e.,  $\nu^\mu$, $\nu_5^\mu$, 
and $\pi^{\mu\nu}$, can be calculated by using the gradient-expansion solution to the CKT equation 
(\ref{eq:CKT-equation-2}), i.e.,
\begin{equation}
\label{eq:f_gradient_expansion}
	f = f_\textrm{eq} - \frac{\tau}{p\cdot u} p\cdot \mathcal{D} f_\textrm{eq} + O(\tau^2\mathcal{D}^2).
\end{equation}
By substituting this distribution function into the definitions in Eqs.~(\ref{eq:current-from-W})--(\ref{eq:Tmunu-from-W}), 
calculating the integrals over the momenta using the formulas in Appendix~\ref{app:integrals}, and 
then extracting the longitudinal and transverse components, we derive the following results up to terms $O(\hbar^2,\hbar\tau\mathcal{D},\tau^2\mathcal{D}^2)$:
\begin{eqnarray}
\label{eq:hydro-coef1}
	\nu^\mu &=& \nu_{\textrm{eq}}^\mu + \frac{\tau}{3} \nabla^\mu n - \tau\dot{u}^\mu n 
	+ \frac{1}{e}\sigma_E E^\mu,
	\\
\label{eq:hydro-coef2}
	\nu_5^\mu &=& \nu_{5,\textrm{eq}}^\mu + \frac{\tau}{3} \nabla^\mu n_5 - \tau\dot{u}^\mu n_5 
	+ \frac{1}{e} \sigma^5_E E^\mu,
	\\
\label{eq:hydro-coef3}
	\pi^{\mu\nu} &=& \frac{8\tau\epsilon}{15} \Delta^{\mu\nu}_{\alpha\beta} (\partial^\alpha u^\beta),
\end{eqnarray}
where $\nu_{\textrm{eq}}^\mu$ and $\nu_{5,\textrm{eq}}^\mu$ are the anomalous nondissipative 
contributions, whose explicit expressions are given in Appendix~\ref{app:integrals}. Note that, by 
definition, $\nabla^\mu \equiv \Delta^{\mu\nu}\partial_\nu = \partial^\mu -u^{\mu} u^{\nu} \partial_\nu$
and $\sigma_E$ is the conventional (positive definite) electrical conductivity that appears in Omh's law,
i.e., $j_{\rm el}^\mu \equiv e \nu^\mu = \sigma_E E^\mu$.
Also, the electric and magnetic fields in the local fluid frame are given by $E^\mu = F^{\mu\nu} u_\nu$ 
and $B^\mu = \tilde F^{\mu\nu} u_\nu$, respectively. 

In the relaxation-time approximation used here, the expressions for the two types of dissipative 
conductivities in Eqs.~(\ref{eq:hydro-coef1}) and (\ref{eq:hydro-coef2}) are given by
\begin{eqnarray}
	\sigma_E &=& \tau e^2\frac{3(\mu^2+\mu_5^2)+\pi^2 T^2 }{9\pi^2\hbar^3},
	\\
\label{eq:sigma_E^5}
	\sigma^5_E &=& \tau e^2 \frac{2\mu\mu_5}{3\pi^2\hbar^3}.
\end{eqnarray}
Furthermore, after taking into account the self-consistency conditions 
(\ref{eq:matching-conditions-n})--(\ref{eq:matching-conditions-h}), 
we arrive at the following first-order hydrodynamic equations: 
\begin{eqnarray}
\label{eq:continuity-1}
	\dot{n} + n \partial_\mu u^\mu + \partial_\mu \nu^\mu &=& 0, \\
\label{eq:continuity-1.5}
	\dot{n}_5 + n_5 \partial_\mu u^\mu + \partial_\mu \nu_5^\mu &=& -\frac{e^2}{2\pi^2\hbar^2} E^\mu B_\mu, \\
\label{eq:continuity-2}
	\dot{\epsilon} + (\epsilon + P) \partial_\mu u^\mu + \partial_\mu h^\mu + u_\mu \dot{h}^\mu - \pi^{\mu\nu} \partial_\mu u_\nu &=& - eE^\mu \nu_\mu, \\
\label{eq:continuity-3}
	(\epsilon + P) \dot{u}^\mu - \nabla^\mu P + h^\alpha \partial_\alpha u^\mu + h^\mu (\partial_\alpha u^\alpha)
	+ \Delta^\mu_\alpha \dot{h}^\alpha 
	+ \Delta^\mu_\alpha \partial_\beta \pi^{\alpha\beta} &=& \epsilon^{\mu\nu\alpha\beta}\nu_\nu u_\alpha eB_\beta + eE^\mu n.
\end{eqnarray}
Finally, by recalling that the electromagnetic fields should be treated as fully dynamical in charged plasmas, 
the above set of equations should be supplemented by the Maxwell equations, i.e., 
\begin{equation}
\label{eq:maxwell}
	\partial_\nu F^{\nu\mu} = en u^\mu + e\nu^\mu - en_{\textrm{bg}} u_{\textrm{bg}}^\mu,
\end{equation}
as well as the Bianchi identity, $\partial_\nu \tilde F^{\nu\mu} = 0$. Note that Eq.~(\ref{eq:maxwell}) 
captures both the Gauss and the Ampere laws. In writing down the corresponding equations, we assumed 
that, in general, the total electric current density may include a nonzero contribution from a static 
nonchiral background, $\rho_{\textrm{bg}}^\mu=en_{\textrm{bg}} u_{\textrm{bg}}^\mu$. Such a contribution 
could play an important role, for example, in cases when a nonzero electric charge of the chiral 
plasma is compensated by a neutralizing background charge of heavy (possibly nonrelativistic) particles. 

By taking into account the constraint for the flow velocity four-vector, $u^\mu u_\mu = 1$, it should be clear
that not all of hydrodynamic equations (\ref{eq:continuity-1})--(\ref{eq:continuity-3}) are truly independent.
Also, by noting that $u^\mu E_\mu = u^\mu B_\mu = 0$, we find that two out of the total eight 
equations (including the Bianchi identity) for the electromagnetic field strength tensor are redundant. 
In fact, as one can check, there are a total of 12 independent equations for 12 variables (i.e., 
six hydrodynamic variables and six components of the electromagnetic field) that govern the hydrodynamic 
evolution of charged chiral plasma.

\section{Equilibrium state}
\label{sec:Equilibrium}

Before addressing the properties of collective modes in a magnetized chiral plasma with nonzero vorticity, 
we should determine the unperturbed (equilibrium) state of the corresponding system. In this section,
we discuss how such a state is defined and what its main properties are. We will assume that the chiral 
charge density $n_{5,\rm{eq}}$ vanishes in equilibrium (i.e., $\mu_5=0$). In the corresponding state, as 
one can see from Eq.~(\ref{eq:sigma_E^5}), the chiral electric separation effect is absent, i.e., $\sigma_E^5=0$,
and several anomalous transport coefficients are trivial, i.e., $\sigma_{\omega}=\sigma_B=\xi_{\omega}=\xi_B=0$,
as is clear from Eqs.~(\ref{eq:anomalous_coeffs-1})--(\ref{eq:anomalous_coeffs-3}) in Appendix~\ref{app:integrals}.

Here it might be appropriate to mention that, despite the absence of an average chiral 
imbalance in equilibrium, it is still appropriate to call the corresponding plasma ``chiral." 
Indeed, local fluctuations of the chirality imbalance can be generically induced by the anomalous 
processes triggered, for example, by collective modes. Also, from a technical viewpoint, the use of 
{\em chiral} hydrodynamics in the description implies that the additional (anomalous) chiral charge 
continuity relation is included in the complete set of equations. The corresponding extra degree of 
freedom can affect the dynamics and modify the properties of collective modes. We might also add 
that, from a conceptual viewpoint, there is no qualitative difference between a long-range hydrodynamic 
fluctuation and a nonzero local average of the chiral imbalance. Formally, this is due to the fact that the 
hydrodynamic description assumes local equilibrium even in the regions with slowly oscillating chiral 
imbalances induced by collective modes.

In order to address the effects of vorticity on hydrodynamic modes, we assume that the background 
vorticity is approximately uniform on distance scales larger than the wavelengths of the modes.
Otherwise, of course, the vortical effects would average to zero. In our study below, we model an 
isolated macroscopic region with approximately uniform vorticity by a uniformly rotating plasma. 
Without loss of generality, we assume that the plasma is confined 
to a cylindrical region of radius $R$ and is uniformly rotating with angular velocity $\Omega$ about the 
$z$ axis. In this study, we will concentrate primarily on the case with $\Omega R\ll 1$. Note, however, 
that many considerations, at least qualitatively, will remain valid even when $\Omega R\lesssim 1$. 
For simplicity, we also assume that the magnetic field $\mathbf{B}$ points along the same $z$ axis. 
While this is clearly not the most general configuration, it is expected to be relevant for applications 
in heavy-ion collisions and the early universe, because the vorticity and magnetic field often tend to
be aligned. 

For the uniformly rotating plasma, the local fluid velocity in the hydrodynamic equilibrium $\bar{u}^\nu$, 
satisfying $\bar{u}^\nu \bar{u}_\nu = 1$, is given by \cite{Florkowski:2018ubm}
\begin{equation}
\label{eq:u_mu}
	\bar{u}^\nu = 
	\gamma  \left(\begin{array}{c} 
		1 \\ -  \Omega y \\  \Omega x \\ 0
	\end{array}\right) ,
\end{equation}
where $\gamma=1/\sqrt{1-\Omega^2r^2}$ and $r \equiv |\mathbf{r}_{\perp}| = \sqrt{x^2+y^2}$. 
Here and below, the notations with bars, such as $\bar{u}^\nu$, represent fields in 
hydrodynamic equilibrium. It should be noted that $\partial_{\mu} \bar{u}^{\mu} =0$ and 
$\Delta^{\mu\nu}_{\alpha\beta} (\partial^\alpha \bar{u}^\beta)=0$. At the same time, the 
radial component of the acceleration is nonzero, $\dot{\bar{u}}^{\mu} = -\gamma^2 \Omega^2 
\mathbf{r}_{\perp}$, as expected for a circular motion. The latter may suggest, however, 
that some dissipative processes are unavoidable in a rotating plasma (even if negligible 
to the linear order in $\Omega$). Indeed, as one can see from Eqs.~(\ref{eq:hydro-coef1}) 
and (\ref{eq:hydro-coef2}), a nonzero $\dot{\bar{u}}^{\mu}$ could potentially trigger dissipative 
currents. As we will see below, this is not necessarily the case because the hydrodynamic 
equilibrium state is radially nonuniform.

As is easy to check, the flow velocity in Eq.~(\ref{eq:u_mu}) is characterized by a nonzero vorticity, i.e.,
\begin{equation}
	\bar\omega^\mu =  \frac{1}{2} \varepsilon^{\mu \nu \alpha \beta} \bar{u}_\nu \partial_\alpha  \bar{u}_\beta
	= \gamma^2 \Omega \delta^{\mu}_{3} .
\end{equation}
It should be mentioned here that, unless stated differently, all explicit expressions for Lorentz
vectors and tensors in the component form will be given from the viewpoint of the laboratory 
frame. In order to obtain these quantities in the comoving frame, 
one would need to perform the corresponding Lorentz transformation $\bar{u}^{\prime\mu}
=\Lambda^{\mu}_{~\nu}  \bar{u}^\nu$ and $\bar{F}^{\prime\mu\nu}=\Lambda^{\mu}_{~\kappa} 
\Lambda^{\nu}_{~\lambda} \bar{F}^{\kappa\lambda}$, where
\begin{equation}
	\Lambda^{\mu}_{~\nu} = 
	\left(\begin{array}{cccc} 
		\gamma & \gamma\Omega y & -\gamma\Omega x & 0 \\ 
		\gamma\Omega y & \frac{x^2+\gamma y^2}{r^2} & \frac{xy(1-\gamma)}{r^2}  & 0 \\ 
		- \gamma\Omega x &  \frac{xy(1-\gamma)}{r^2}  & \frac{\gamma x^2+y^2}{r^2}  & 0\\ 
		0 & 0 & 0 & 1
	\end{array}\right) .
\end{equation}
For example, for the fluid flow velocity, we obtain $ \bar{u}^{\prime\mu}=(1,0,0,0)$, 
as expected in the local comoving frame. 

In the presence of a magnetic field, the definition of an equilibrium state of a charged rotating plasma 
is far from trivial. For example, a naive assumption that, in the lab frame, there is only a nonzero 
magnetic field pointing in the $z$ direction is not self-consistent. Indeed, for such a configuration, there 
would be nonzero electric fields present in the comoving frame. Since such electric fields would drive 
nonvanishing dissipative currents [see Eqs.~(\ref{eq:hydro-coef1}) and (\ref{eq:hydro-coef2})], it would 
imply that the thermodynamic equilibrium is not reached in the local frame of the fluid. 

\begin{figure}[t]
		\includegraphics[width=0.4\textwidth]{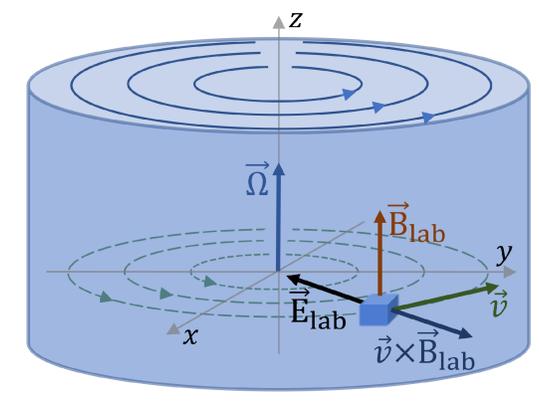}
		\caption{A charged element in a rotating plasma and the 
		local electromagnetic fields. Note that the electric and Lorentz forces on a 
		local element of plasma are equal in magnitude and opposite in direction.}
	\label{fig:equilibrium_state}
	\end{figure}

In order to construct the global equilibrium state of the magnetized rotating plasma, therefore, we 
require that the electric fields vanish in the local fluid frame, i.e., $\bar{E}^\mu \equiv \bar{F}^{\mu\nu} \bar{u}_\nu=0$. 
As for the magnetic field in the same local frame $\bar{B}^\mu \equiv \tilde{F}^{\mu\nu} \bar{u}_\nu$, 
we assume only that it points in the $z$ direction, i.e., $\bar{B}^\mu = B\delta^{\mu}_{3}$, and its magnitude 
may have a nontrivial dependence on the cylindrical radius coordinate, i.e., $B\equiv B(r)$. Then, the 
corresponding electromagnetic field stress tensor in the lab frame takes the following form:
\begin{equation}
\label{eq:Fmunu}
\bar{F}^{\mu\nu}  = \epsilon^{\mu\nu\alpha\beta} \bar{u}_\alpha \bar{B}_\beta 
			+ \bar{E}^\mu \bar{u}^\nu - \bar{u}^\mu \bar{E}^\nu = \left(\begin{array}{cccc} 
		0 & \gamma B \Omega x & \gamma B \Omega y & 0 \\ 
		- \gamma B \Omega  x  & 0 & -\gamma B  & 0 \\ 
		- \gamma B \Omega  y &  \gamma B  & 0 & 0\\ 
		0 & 0 & 0 & 0
	\end{array}\right).
\end{equation}
As we see, this includes not only a magnetic field in the $z$ direction, 
$\mathbf{B}_{\rm lab} =\gamma B\hat{\mathbf{z}}$, 
but also an electric field in the radial direction, i.e., 
$\mathbf{E}_{\rm lab}=-\gamma B \Omega\mathbf{r}_\perp$.
Alternatively, this can be viewed as the consequence of Ohm's law for an ideal 
plasma, $\mathbf{E}_\textrm{lab} = - \mathbf{v}\times \mathbf{B}_{\rm lab}$.
Note that such a configuration in the lab frame is possible because the electric force on 
a local element of the charged fluid is exactly compensated by the Lorentz force, see 
Fig.~\ref{fig:equilibrium_state}. In order to be consistent with Ampere's law, as we will argue 
below, the magnetic field $B$ should additionally have a very specific dependence on 
the radial coordinate. 

For the configuration in Eq.~(\ref{eq:Fmunu}), the Bianchi identity
is satisfied identically, while the Maxwell equations take the following explicit form:
\begin{equation}
\label{eq:Maxwell_equilb}
	-\delta^{\mu}_{0} \Omega \left[ 2 \gamma B + r \partial_{r } (\gamma B) \right]
	+\left(\delta^{\mu}_{1} \frac{y}{r } 
	-\delta^{\mu}_{2} \frac{x}{r } \right) \partial_{r }  (\gamma B ) 
	= en_{\textrm{eq}} \bar{u}^\mu - en_{\textrm{bg}} u_{\textrm{bg}}^\mu .
\end{equation}
Here we took into account that $e\nu_{\textrm{eq}}^\mu =0$, which is indeed the case when $\mu_5=0$. 
Let us note in passing that major modifications would be needed in the analysis if one allows for a 
nonzero $\mu_5$. Indeed, in order to enforce the hydrodynamic equilibrium and the absence of 
dissipative currents at $\mu_5 \neq 0$, additional nonzero components of the electromagnetic field
would be required.

We assume that the background charge is at rest in the lab frame, i.e., 
$u_{\textrm{bg},1}^\mu =(1,0,0,0)$. (It might also be interesting to consider a comoving 
background with $u_{\textrm{bg},2}^\mu =\bar{u}^\mu$, but we will not investigate such 
a possibility in this paper.) As is easy to check, the Maxwell equations require that the magnetic 
field and the charge density of the plasma have the following dependence on the radial coordinate:
\begin{eqnarray}
\label{eq:B_vs_r_perp}
   B(r) &=& \gamma \left(B_0- \frac{1}{2} en_{\textrm{bg}} \Omega r^2\right)
 \simeq  B_0 - \frac{1}{2} en_{\textrm{bg}} \Omega r^2 +O\left(B_0 r^2\Omega^2\right) ,\\
\label{eq:n_vs_r_perp}
   en_{\textrm{eq}} (r) &=& \gamma^3 \left(en_{\textrm{bg}} -2B_0 \Omega \right)
   \simeq  en_{\textrm{bg}} - 2 B_0 \Omega  +O\left(en_{\textrm{bg}} r^2 \Omega^2\right),
\end{eqnarray}
where $B_0$ is the value of the magnetic field on the rotation axis (i.e., at $r=0$). An interesting feature 
of this solution is that $\frac{1}{3} \nabla^\mu n_{\textrm{eq}}(r) - \dot{\bar{u}}^\mu n_{\textrm{eq}}(r) = 0$
to all orders in $\Omega$. This means, in particular, that the dissipative part of the fermion number
(electric) current in 
Eq.~(\ref{eq:hydro-coef1}) vanishes exactly. This is quite amazing considering that the solution was 
obtained by solving the Maxwell equations (\ref{eq:Maxwell_equilb}) without any explicit considerations 
of dissipative effects. 

We should remark that, in the absence of a background charge (i.e., $n_{\textrm{bg}}=0$), 
the solutions in Eqs.~(\ref{eq:B_vs_r_perp}) and (\ref{eq:n_vs_r_perp}) appear to agree with the 
vortexlike solutions found in Ref.~\cite{Florkowski:2018ubm}. The dependence of the plasma 
density on the $\gamma$ factor in Eq.~(\ref{eq:n_vs_r_perp}) is also consistent with the findings 
in Ref.~\cite{Becattini:2009wh}, stating that the local chemical potentials and temperature 
should be linear in $\gamma$: $\bar{\mu}(r)=\gamma \mu_0$, $\bar{\mu}_5(r)=\gamma \mu_{5,0}$, and 
$\bar{T}(r)=\gamma T_0$, where $\mu_0$, $\mu_{5,0}$, and $T_0$ are the values of the corresponding 
parameters on the rotation axis. Indeed, in the case of a charged plasma made of massless particles, 
such a linear dependence automatically leads to the scaling law in Eq.~(\ref{eq:n_vs_r_perp}).
(It is not clear, though, whether similar arguments can be generalized for a plasma made of massive 
particles.)

As we see from the solutions in Eqs.~(\ref{eq:B_vs_r_perp}) and (\ref{eq:n_vs_r_perp}), the 
Maxwell equations require that the local charge density is nonzero  in a rotating plasma
(e.g., to the leading order in vorticity, $en_{\textrm{eq}} -en_{\textrm{bg}}\simeq - 2 B_0 \Omega$). 
From a physics viewpoint, one might interpret this as a charge separation induced by the uniform 
rotation. A nonzero electric charge in the bulk is achieved by pushing the compensating charge of 
the opposite sign out to the cylindrical boundary of the system. Interestingly, the sign of the induced 
charge in the bulk is determined by the relative orientation of the magnetic field and vorticity and, 
thus, it can easily be flipped, for example, by changing the direction of the magnetic field.

It should be noted that a charge separation in the rotating plasma is consistent with the fact 
that there is a nonzero electric field in the radial direction, $\mathbf{E}_{\rm lab}=-\gamma B 
\Omega\mathbf{r}_\perp$. As mentioned earlier, the electric force of such 
a field on a moving element of the charged plasma is exactly compensated by the Lorentz 
force from the magnetic field, $\mathbf{B}_{\rm lab} =\gamma B\hat{\mathbf{z}}$; see 
Fig.~\ref{fig:equilibrium_state}. Additionally, in agreement with Ampere's law, the circular 
currents of the plasma result in a total magnetic field that depends on the radial coordinate,
see Eq.~(\ref{eq:B_vs_r_perp}). In the presence of a background charge, moreover, the nontrivial 
dependence of the field on the radial coordinate appears already at the linear order in $\Omega$, 
i.e., $B(r) \simeq B_0 - \frac{1}{2} en_{\textrm{bg}} \Omega r^2$.

\section{Linearized equations of chiral hydrodynamics}
\label{sec:Linearized}

In order to set the stage for a systematic study of hydrodynamic collective modes in a 
rotating chiral plasma, here we will discuss how to derive the linearized equations for small 
perturbations of the hydrodynamic quantities around their equilibrium values. In the derivation, 
it is beneficial to take into account the symmetry of the unperturbed state with respect to 
time translations, translations in the spatial $z$ direction, as well as the rotational symmetry 
about the $z$ axis. (Recall that both the background magnetic field and the vorticity are 
parallel to the $z$ axis.) 

By taking into account the cylindrical symmetry in the problem, it is convenient to write down 
the general wavelike perturbations of (pseudo)scalar and (pseudo)vector quantities 
in the following form:
\begin{eqnarray}
\label{eq:perturbations-1}
	\delta s(x) &=& e^{-i k_{0} t + i k_{z} z + im\theta} \delta s(r),	\\
\label{eq:perturbations-2}
	\delta v^3(x) &=& e^{-i k_{0} t + i k_{z} z + im\theta} \delta v^3(r),	\\
\label{eq:perturbations-3}  
	\delta v_\pm(x) &=& e^{-i k_{0} t + i k_{z} z + i(m\pm 1)\theta}   \delta v_\pm(r) ,
\end{eqnarray}
where $s$ is a placeholder for all scalar and pseudoscalar parameters (i.e.,  $\mu$, $\mu_5$, or $T$). 
Similarly, the generic notations $v^3$ and $v_\pm$ stand for the longitudinal (i.e., $u^3$, $B^3$, or $E^3$) 
and transverse (i.e., $u_\pm$, $B_\pm$, or $E_\pm$) components of the vector quantities. Note that 
the plus and minus components are defined as follows: $v_{\pm}=\frac{1}{2}\left(v^{1} \pm i v^{2}\right)$. 
Such a separation of the longitudinal and transverse components, as well as their dependence on the 
cylindrical radius $r =\sqrt{x^2+y^2}$ and the polar angle $\theta = \textrm{arctan}(y/x)$, follow from the 
requirement that the corresponding perturbations are the eigenstates of the angular momentum operator 
$\hat{J}_z$ with the eigenvalues $m\in\mathbb{Z}$. [Note that $\hat{J}_z= - i \partial_\theta $ for scalar 
and pseudoscalar fields and $\hat{J}_z= - i\partial_\theta +S_{z}$ with $(S_{z})^{i}_{~j} = - i\varepsilon^{ij3}$
for the vector fields.] In agreement with the remaining translational symmetry, the only well-defined
components of the wave vector are the timelike and longitudinal ones, i.e., $k_{0}$ and $k_{z}$, 
respectively.

By taking into account the constraints for the flow velocity and the electromagnetic field (i.e., 
$u^\mu u_\mu = 1$ and $B^\mu u_\mu = E^\mu u_\mu = 0$, respectively) it is clear that only 
three out of the total four components in each four-vector are independent. Without loss of generality, 
we assume that the spatial components are the independent ones. Then, as is easy to check, the deviations 
of the zeroth components from their equilibrium values are not independent. They are given by 
\begin{eqnarray}
\label{eq:perturb-u0}
	\delta u^0 &=& \Omega \left[ x \delta u^{2}(x) - y \delta u^{1}(x) \right] 
			= i \Omega r \left[ e^{i\theta}\delta u_{-}(x) - e^{-i\theta}\delta u_{+}(x)\right] , \\
\label{eq:perturb-E0}
	\delta E^0 &=& \Omega \left[ x \delta E^{2}(x) - y \delta E^{1}(x)\right] 
			= i \Omega r  \left[ e^{i\theta}\delta E_{-}(x) - e^{-i\theta}\delta E_{+}(x)\right] , \\
\label{eq:perturb-B0}
	\delta B^0 &=&\Omega \left[ x \delta B^{2}(x) - y \delta B^{1}(x)\right] + B\delta u^{3}(x) 
			= i \Omega r  \left[ e^{i\theta}\delta B_{-}(x) - e^{-i\theta}\delta B_{+}(x)\right] 
			+ B \delta u^{3} (x) .
\end{eqnarray}
It should be noted, that the general form of the perturbation to the electromagnetic field strength 
tensor in the lab frame can be conveniently written in terms of the perturbations of the electric and
magnetic fields as follows:
\begin{eqnarray}
	\delta F^{\mu\nu} &=& \epsilon^{\mu\nu\alpha\beta} \left(\delta u_\alpha \bar{B}_\beta 
				+  \bar{u}_\alpha \delta B_\beta \right) + \delta E^\mu \bar{u}^\nu - \bar{u}^\mu \delta E^\nu ,\\	
	\delta \tilde F^{\mu\nu} &=& \epsilon^{\mu\nu\alpha\beta} \delta E_\alpha \bar{u}_\beta 
				+ \delta B^\mu \bar{u}^\nu - \bar{u}^\mu \delta B^\nu  
				 + \bar{B}^\mu \delta u^\nu - \delta u^\mu \bar{B}^\nu,
\end{eqnarray}
where we took into account that the electric field is absent in the unperturbed state. By making use of 
these relations and assuming that all perturbations are small, it is straightforward to obtain a linearized 
system of hydrodynamic equations from Eqs.~(\ref{eq:continuity-1})--(\ref{eq:maxwell}). The 
corresponding linearized equations are presented in Appendix~\ref{app:Full_system}. Note that no 
explicit dependence on $t$, $z$, and $\theta$ coordinates is present in those equations because the 
common exponent $e^{-i k_{0} t + i k_{z} z+im\theta}$, containing such a dependence, was factored out. 

In order to solve the hydrodynamic equations for collective modes, we should also 
specify the boundary conditions for the fields. For simplicity, we will assume that the perturbations 
of the scalar, pseudoscalar and longitudinal vector perturbations vanish at the boundary, i.e., 
$\delta s(R) = 0$ and $\delta v^3(R) = 0$, where $R$ is the cylindrical radius of the system. 
[Note, however, that $\delta v_\pm(R) = 0$ cannot be enforced at the same time.] It should 
be clear, however, that the dispersion relations for most of the bulk modes will not be very sensitive 
to the choice of the boundary conditions, unless their transverse wave vectors are very small, 
$k_\perp \simeq 1/R$ (the exact definition of $k_\perp$ will be specified later).

\subsection{The linearized equations at vanishing vorticity}

Before discussing any solutions in the most general case with a nonzero vorticity and magnetic field,
let us first consider a simple setup without rotation, $\Omega =0$, but with the effects of dynamical 
electromagnetism fully accounted for. As will be clear, in our analysis, we also include all effects associated 
with the dynamical vorticity induced by collective modes themselves.

The linearized system of the hydrodynamic equations takes the following explicit form:
\begin{eqnarray}
\label{eq:system-simple-start}
	u^\mu\partial_\mu\delta n + n\partial_\mu\delta u^\mu + B^\mu\partial_\mu \delta\sigma_B + \frac{\tau}{3}\nabla_\mu\nabla^\mu\delta n - \tau nu^\nu\partial_\mu\partial_\nu\delta u^\mu + \frac{1}{e} \sigma_E \partial_\mu\delta E^\mu &=& 0,
	\\ 
\label{eq:system-simple-start001}
	u^\mu\partial_\mu\delta n_5 + \sigma_B^5 \partial_\mu \delta B^\mu + B^\mu\partial_\mu\delta\sigma_B^5 + \frac{\tau}{3}\nabla_\mu\nabla^\mu\delta n_5  +\frac{e^2}{2\pi^2\hbar^2} (B\cdot\delta E) &=& 0,
	\\
	u^\mu\partial_\mu\delta \epsilon + (\epsilon + P)\partial_\mu\delta u^\mu + B^\mu\partial_\mu\delta\xi_B &=& 0,
	\\
	(\epsilon + P) u^\nu\partial_\nu\delta u^\mu - \nabla^\mu\delta P + B^\mu u^\nu\partial_\nu\delta\xi_B + \frac{8\tau\epsilon}{15} \Delta^{\mu\nu}_{\alpha\beta} (\partial_\nu\partial^\alpha \delta u^\beta)  -en\delta E^\mu &&
	\nonumber\\
	- \epsilon^{\mu\nu\alpha\beta}\left(\frac{\tau}{3}\nabla_\nu\delta n - \tau n u^\phi\partial_\phi\delta u_\nu 
	+ \frac{1}{e}\sigma_E \delta E_\nu \right) u_\alpha eB_\beta&=& 0,
\label{eq:system-simple-end-hydro}
\end{eqnarray}
which should also be supplemented by the Maxwell equations
\begin{eqnarray}
\label{Maxwell111}
	\epsilon^{\mu\nu\alpha\beta}u_\nu\partial_\alpha\delta E_\beta + u^\nu\partial_\nu\delta B^\mu + B^\mu \partial_\nu\delta u^\nu - B^\nu\partial_\nu\delta u^\mu - u^\mu\partial_\nu \delta B^\nu &=& 0,
	\\
	\epsilon^{\mu\nu\alpha\beta}(u_\nu\partial_\alpha\delta B_\beta - B_\nu\partial_\alpha\delta u_\beta) + u^\mu\partial_\nu\delta E^\nu - u^\nu\partial_\nu\delta E^\mu - en\delta u^\mu - eu^\mu\delta n &&
	\nonumber\\
	\qquad
	 - e\delta\sigma_B B^\mu - \frac{e\tau}{3}\nabla^\mu \delta n + e\tau n u^\phi\partial_\phi\delta u^\mu - \sigma_E\delta E^\mu &=& 0.
\label{eq:system-simple-end}
\end{eqnarray}
The variations of the fermion number density, as well as other functions (e.g., $\delta\epsilon$ and $\delta\sigma_B$) 
are assumed to be of the most general form, i.e.,  $\delta n = \frac{\partial n}{\partial\mu} \delta\mu (x)
+ \frac{\partial n}{\partial\mu_5} \delta\mu_5 (x) + \frac{\partial n}{\partial T} \delta T(x)$. They are space and 
time dependent, although such a dependence may not always be shown explicitly, e.g., $\delta n\equiv \delta n(x)$ and
$\delta n_5\equiv\delta n_5(x)$.

In the case of vanishing vorticity, the independent wavelike solutions of the hydrodynamic equations 
can be conveniently expressed in terms of the cylindrical harmonics. The corresponding solutions 
for the local perturbations are given by Eqs.~(\ref{eq:perturbations-1})--(\ref{eq:perturbations-3}), 
with radial parts of the functions given by
\begin{eqnarray}
\label{eq:ansatz-1}
	\delta s(r) &=& \delta s\, J_m(k_\perp r), 
	\quad\mbox{for}\quad
	s = \mu,\mu_5,T, \\
\label{eq:ansatz-2}
	\delta v^3(r) &=& \delta v^3\, J_m(k_\perp r), 
	\quad\mbox{for}\quad
	v^3 = u^3,B^3,E^3, \\
\label{eq:ansatz-3}
	\delta v_\pm(r) &=& \delta v_\pm\, J_{m\pm 1}(k_\perp r),
	\quad\mbox{for}\quad
	v_\pm = u_\pm, B_\pm, E_\pm.
\end{eqnarray}
Here $J_m(k_\perp r)$ is the Bessel function of the first kind and parameter $k_\perp$ is an analogue 
of the transverse wave vector for a system with cylindrical symmetry.

The linearized system (\ref{eq:system-simple-start})--(\ref{eq:system-simple-end}) has the general 
structure $\hat M \delta\vec{f} = 0$, where $\hat M$ is a $12\times 12$ matrix differential operator and 
$\delta\vec{f}$ is a vector consisting of all independent plasma and EM field perturbations, i.e., $\delta\vec{f} 
= (\delta\mu, \delta\mu_5, \delta T, \delta \mathbf{u}, \delta \mathbf{E}, \delta \mathbf{B})^T$. 
By making use of the ansatz in Eqs.~(\ref{eq:perturbations-1})--(\ref{eq:perturbations-3}) 
with the radial dependence in Eqs.~(\ref{eq:ansatz-1})--(\ref{eq:ansatz-3}),
it is easy to check that all coordinate dependence factorizes and the problem reduces to 
a set of homogeneous linear equations, $M \delta\vec{f} = 0$, where $M$ is a $12\times 12$ algebraic 
matrix. For the system to have a nontrivial solution, the characteristic equation should be 
satisfied, i.e., $\mbox{det}(M) =0$. In essence, the latter is a polynomial equation for the 
frequencies of collective modes. The roots for $k_0$ define dispersion relations of 
hydrodynamic modes. In general, the corresponding frequencies (energies) are functions 
of the wave vectors $k_\parallel$ and $k_\perp$, as well as the eigenvalue $m$ of the angular 
momentum operator. The associated nontrivial solutions for $\delta\vec{f}$ (``eigenvectors") specify 
the nature of the collective modes.

It is instructive to note that the above general procedure for obtaining the dispersion relations of 
hydrodynamic modes could easily be adjusted to take into account any boundary conditions 
consistent with the cylindrical symmetry. As we mentioned earlier, we will assume that the 
perturbations vanish at the cylindrical boundary of the system. Such boundary conditions are
satisfied automatically when the values of the transverse wave vector are restricted to take only 
the following discrete values: $k_\perp^{(i)} = \alpha_{m, i}/R$, where $\alpha_{m, i}$ (with $i = 1, 2, ...$) 
is the $i$th zero of the Bessel function $J_m(z)$. By making use of the asymptotic form for the Bessel 
function, we can derive the following approximate expression for the large transverse wave vectors: 
$k_\perp^{(i)} \simeq (i+m/2-1/4)\pi/R$ for $i\gg m$. By imposing the periodic boundary conditions 
in the $z$ direction (with period $L$), the longitudinal wave 
vector is discretized, i.e., $k_z^{(j)}= 2\pi j/L$, where $j$ is an integer. When the system is large, the discretized 
wave vectors of both types are closely located and, thus, become almost indistinguishable from a 
continuum. In such a case, the discretization plays little role and could be ignored.

\subsection{The linearized equations at nonzero vorticity}

In the general case of a rotating plasma, the self-consistent analysis of hydrodynamic modes 
becomes much more complicated. One of the primary reasons for this complication is the nontrivial radial 
dependence of the magnetic field and density in the unperturbed state of a uniformly rotating plasma;
see Eqs.~(\ref{eq:B_vs_r_perp}) and (\ref{eq:n_vs_r_perp}). In this case, the radial parts of  
local perturbations in Eqs.~(\ref{eq:ansatz-1})--(\ref{eq:ansatz-3}) can be written in the form of
Fourier-Bessel series:
\begin{eqnarray}
\label{eq:ansatz-series-1}
	\delta s(r) &=&\sum_{i=1}^\infty \delta s^{(i)}\, J_m(k_\perp^{(i)} r), 
	\quad\mbox{for}\quad
	s = \mu,\mu_5,T,
	\\
\label{eq:ansatz-series-2}
	\delta v^3(r) &=& \sum_{i=1}^\infty \delta v^3\, J_m(k_\perp^{(i)} r), 
	\quad\mbox{for}\quad
	v^3 = u^3,B^3,E^3, \\
\label{eq:ansatz-series-3}
	\delta v_\pm(r) &=& \sum_{i=1}^\infty \delta v_\pm^{(i)}\, J_{m\pm 1}(k_\perp^{(i)} r),
	\quad\mbox{for}\quad
	v_\pm = u_\pm, B_\pm, E_\pm,
\end{eqnarray}
where $k_\perp^{(i)}= \alpha_{m, i}/R$ is the $i$th discretized value of the transverse wave vector,
introduced in the previous subsection. The set of Bessel eigenfunctions used in the series above is 
complete and orthogonal \cite{completness1,completness2}. The orthogonality condition and other useful properties 
of the Bessel eigenfunctions are discussed in Appendix~\ref{app:Bessel}.

By substituting perturbations~(\ref{eq:ansatz-series-1})--(\ref{eq:ansatz-series-3}) 
into the complete set of hydrodynamic equations (see Appendix~\ref{app:Full_system}) and projecting 
the results onto the Bessel functions with different $k_\perp^{(i)}$, we arrive at an algebraic system 
of equations with the following block-matrix form:
\begin{equation} 
\label{eq:block-matrix-equation}
	\left(\begin{array}{ccc}
		\begin{array}{c}
				M^{(11)} 
		\end{array}
		& 
		\begin{array}{c}
				M^{(21)} 
		\end{array}
		& \dots
		\\
		\begin{array}{c}
				M^{(12)} 
		\end{array}
		&
		\begin{array}{c}
				M^{(22)} 
		\end{array}
		& \dots
		\\
		\vdots & \vdots & \ddots
	\end{array}\right)
	\cdot
	\left(\begin{array}{c}
		\begin{array}{c}
				\delta\vec{f}^{(1)} 
		\\
				\delta\vec{f}^{(2)}
		\end{array}
		\\
		\vdots
	\end{array}\right)
	=
	\left(\begin{array}{c}
		\vec{0} \\ \vec{0} \\ 
		\vdots
	\end{array}\right),
\end{equation}
where $M^{(ij)}$ are $12\times 12$ matrices made of the $i$th set of coefficient functions projected onto the $j$th 
Bessel eigenfunctions, and $\delta\vec{f}^{(i)} = (\delta\mu^{(i)} , \delta\mu_5^{(i)} , \delta T^{(i)} , \delta \mathbf{u}^{(i)} , 
\delta \mathbf{E}^{(i)} , \delta \mathbf{B}^{(i)} )^T$. Formally, Eq.~(\ref{eq:block-matrix-equation}) is an infinite 
system of equations. By noting, however, that the hydrodynamic description is limited to a finite range of low 
energies and momenta, the system can be truncated in a self-consistent way. In general, it is sufficient to limit 
the sum in Eqs.~(\ref{eq:ansatz-series-1})--(\ref{eq:ansatz-series-3}) to values of the transverse wave vector
$k_\perp^{(i)}$ that are less than $1/l_{\rm mfp}$, where $l_{\rm mfp}$ is the mean free path of particles. In 
practice, though, when focusing on the low-energy part of the spectrum and/or studying the limit of small 
vorticity, the truncation could be even more restrictive. 

Before proceeding further with the analysis, it is instructive to discuss the dependence of the matrix blocks 
in Eq.~(\ref{eq:block-matrix-equation}) on the angular velocity $\Omega$. As is clear from the discussion 
in the previous subsection, in the absence of vorticity (i.e., at $\Omega=0$), all off-diagonal blocks in 
Eq.~(\ref{eq:block-matrix-equation}) vanish. Indeed, in such a limit, hydrodynamic modes are characterized by 
well-defined values of $k_\perp^{(i)}$. Also, the energies of the corresponding modes are determined by the 
roots of the characteristic equations $\mbox{det}(M^{(ii)})=0$, where $M^{(ii)}$ are the diagonal blocks 
associated with specific values $k_\perp^{(i)}$. 

At nonzero vorticity, the off-diagonal blocks in Eq.~(\ref{eq:block-matrix-equation}) do not vanish any more 
and, as a result, all hydrodynamic modes become nontrivial mixtures of partial waves with different values 
of $k_\perp^{(i)}$. In principle, the corresponding spectrum could be obtained by solving  
Eq.~(\ref{eq:block-matrix-equation}) using numerical methods. While such an approach is straightforward 
conceptually, it is rather tedious and is beyond the scope of this study. Instead, in the rest of this paper, 
we will investigate the limit of small, but nonzero vorticity. 

In order to determine the spectrum of collective modes at small $\Omega$, we will solve the characteristic 
equations $\mbox{det}(M)=0$ to leading order in $\Omega$. While this may seem to be a very strong 
limitation, it should be noted that even rather optimistic estimates of vorticity in heavy-ion collisions are 
not that large in relative terms \cite{Becattini:2015ska,Jiang:2016woz,Deng:2016gyh}. 
As is easy to check, in the limit of small $\Omega$, 
nontrivial corrections to the off-diagonal blocks in Eq.~(\ref{eq:block-matrix-equation}) start from the {\em linear} 
order terms in $\Omega$. In the calculation of $\mbox{det}(M)$, therefore, such off-diagonal matrices 
contribute only starting at the {\em quadratic} order in $\Omega$. This means, in particular, that all linear 
corrections to the dispersion relations of hydrodynamic modes are determined completely by the linear 
corrections to the diagonal blocks. The corresponding characteristic equations are given by 
$\mbox{det}(M^{(ii)}+\Omega M_1^{(ii)})=0$, where $M_1^{(ii)} \equiv (\partial M^{(ii)}/\partial \Omega)|_{\Omega=0}$. 

From the above general structure of the approximate characteristic equations, it should also be clear that, to 
linear order in $\Omega$, the hydrodynamic modes can be unambiguously labeled by well-defined values of 
the transverse wave vector $k_\perp^{(i)}$. In other words, while the dispersion relations of the modes are modified 
by the vorticity, their classification remains the same as in the $\Omega=0$ case. Of course, this is hardly 
surprising and should have been expected since, in essence, we used a perturbation theory with $\Omega$ 
playing the role of a small parameter. In this connection, we should add though that, starting already from the 
quadratic order in $\Omega$, the off-diagonal blocks in Eq.~(\ref{eq:block-matrix-equation}) are non-negligible and a substantial 
mixing of partial waves with different $k_\perp^{(i)}$ will be unavoidable.

\section{Hydrodynamic modes in high temperature plasma}
\label{sec:Modes-highT}

In this section, we study the spectrum of hydrodynamic modes in a chiral rotating plasma at high 
temperature, i.e., $T\gg \mu$, or, in other words, we assume that the fermion number chemical 
potential is small compared to the temperature. As is clear, such a regime is most suitable for
describing hot plasmas in the early universe and in heavy-ion collisions. The opposite limit, i.e., 
$T\ll \mu$, will be addressed in the next section. 

Before proceeding to the technical part of the study, it is instructive to discuss the general validity 
of the hydrodynamic approach and the hierarchy of various length scales in the problem at hand. 
The shortest length scale of relevance is the de Broglie wavelength for chiral particles $l_{d} 
\simeq \hbar/T$ (at high density, it will be replaced by $l_{d} \simeq \hbar/\mu$). Clearly, the 
hydrodynamic description cannot work on such short microscopic distances. In fact, it becomes 
relevant only on the scales much larger than the particle mean free path, i.e., $l_{\rm mfp} \simeq 
\tau$ (recall that $c=1$ in the units used here), where the definition of local equilibrium could be 
meaningful. In a background magnetic field, there is an additional scale defined 
by the magnetic length $l_B = \sqrt{\hbar/eB}$. We will assume that the field is weak in the sense 
that $\sqrt{\hbar eB}\ll T$, which 
is usually the case in most applications. This implies that $l_{d} \ll l_B$. The magnetic length could, 
however, be comparable to the mean free path. In fact, the hierarchy between the two scales could
be used to distinguish the regime of very weak fields, i.e., $l_B\gtrsim l_{\rm mfp}$, from that of  
moderately strong fields, i.e., $l_B\lesssim l_{\rm mfp}$.

When discussing hydrodynamic modes, we will have to deal with yet another window of 
length scales, defined by the wavelengths of the modes, i.e., $\lambda_k\simeq 1/k$, 
where $k$ is the corresponding wave vector. Clearly, the hydrodynamic description for such modes
makes sense only if $\lambda_k \gg l_{\rm mfp}$ (or, equivalently, $k \ll 1/l_{\rm mfp}$). In 
a finite system, at the same time, the maximum wavelength is limited by the size of the system 
itself, $\lambda_k \lesssim R$. Finally, the size of a uniformly rotating relativistic plasma is 
limited by the scale of $\Omega^{-1}$. Therefore, in the analysis of collective modes 
below, we will assume the following hierarchy of scales: $l_{d} \ll l_B \lesssim l_{\rm mfp} \ll 
\lambda_k \lesssim R \ll \Omega^{-1}$. This hierarchy will also be used in the derivation 
of analytical results. In order to simplify the task of keeping track of different scales, it will be 
convenient to use the following values and ranges of dimensionless parameters:  
\begin{equation}
\label{eq:scales_hot_plasma}
	l_{\rm mfp} \Omega  \simeq \xi^2 , 
	\qquad
	\xi^{3/2}\simeq\frac{l_{\rm mfp}}{R}\lesssim k l_{\rm mfp} \lesssim\xi^{1/2},
	\qquad
	\frac{l_{\rm mfp}}{l_B} \simeq \xi^{-1/4} ,
	\qquad
	\frac{l_{\rm mfp}}{l_{d}} \simeq \xi^{-1} ,
\end{equation}
where we introduced a small parameter $\xi \simeq 10^{-2}$ in order to easily separate all relevant 
scales in the problem. While concentrating on the high-temperature regime here, it might 
be interesting to see how the effects of a small chemical potential start showing up in the spectrum 
of collective modes. Therefore, we also include a nonzero $\mu$, but assume that its value is very 
small, e.g., $ |\mu| \simeq \xi^2 T$ (or, equivalently, $|\mu| l_{d} \simeq \xi^2 \hbar$).

\subsection{Charged plasma at $\Omega=0$}

Before addressing hydrodynamic modes at nonzero vorticity, it is instructive to set up the stage by first 
discussing the benchmark results at $\Omega=0$. As expected in the hydrodynamic regime, there are 
generically two very different types of modes: propagating and diffusive. The dispersion relations of the 
former ones have nonzero real parts and relatively small imaginary parts. The diffusive modes, on the 
other hand, have either no real parts at all, or the imaginary parts much larger than real ones. 

By solving the linearized equations (\ref{eq:system-simple-start})--(\ref{eq:system-simple-end}), we find 
that there are two kinds of propagating modes at $\Omega=0$, namely, a sound wave with the 
dispersion relation given by
\begin{equation}
\label{eq:s-wave-00}
k_0 = \frac{s_{e}k}{\sqrt{3}} - \frac{2}{15}i\tau k^2 ,
\end{equation}
and an Alfv\'en wave with the dispersion relation approximately given by
\begin{equation}
\label{eq:Alfven-wave-00}
k_0^{(\pm)} = s_{e}\frac{3\sqrt{5} B\hbar^{3/2} k_z}{\sqrt{7}\pi T^2}
\left( 1 \pm \frac{\sqrt{5} e\mu}{2\sqrt{7}\pi\hbar^{3/2} k} \right) - \frac{1}{10}i\tau k^2 .
\end{equation}
In both cases, $k=\sqrt{k_z^2+k_\perp^2}$ and $s_{e}=\pm1$.  As is clear, both choices of $s_{e}$ (i.e., 
the overall sign in front of the real part of the energy) correspond to the same mode. In most cases, 
the signs $s_{e}$ could be associated simply with different directions of propagation. As is easy to 
check, in fact, this is the case for the Alfv\'en waves in Eq.~(\ref{eq:Alfven-wave-00}).

The sound wave in Eq.~(\ref{eq:s-wave-00}) describes the propagation of longitudinal elastic deformations 
in plasma. The propagation of such a wave does not induce any local oscillations of the electric charge 
and, as a result, there are no dynamical electromagnetic fields induced. Also, as expected for the ultrarelativistic 
matter, the speed of sound $c_s$ is determined by its compressibility, $c_s^2 = \partial P/\partial\rho = 1/3$. 
(As we mentioned earlier, in a more realistic case of a strongly interacting quark-gluon plasma, 
the value of $c_s^2$ is expected to be somewhat smaller than $1/3$ \cite{Bazavov:2014pvz}, but there is 
no reason to expect that the nature of the corresponding sound mode will change qualitatively.)

A few words are in order about the Alfv\'en waves with the dispersion relations given by Eq.~(\ref{eq:Alfven-wave-00}). 
These are magnetohydrodynamic modes with two possible circular polarizations: the left-handed one with 
$\delta u_+ \ll \delta u_-$ and the right-handed one with $\delta u_+ \gg \delta u_-$. From the viewpoint
of the fluid flow oscillations, these are transverse modes. The energies of the 
corresponding two branches of waves differ slightly because the term with the chemical potential $\mu$ comes
with opposite signs. In the limit $\mu\to 0$, the speed of propagation of these waves is the same for both 
polarizations. As is easy to check by neglecting the dissipative effects, the expression for the speed can also  
be written as $v_A= B/\sqrt{\epsilon_{\textrm{eq}}+P_{\textrm{eq}}}$,  which is the standard expression
for the Alfv\'en waves in a relativistic plasma \cite{Harris}. It might be appropriate to mention that the propagation of 
Alfv\'en waves is accompanied by the fluid flow oscillations with nonzero local dynamical vorticity.

\begin{figure}[t]
    \includegraphics[width=0.32\textwidth]{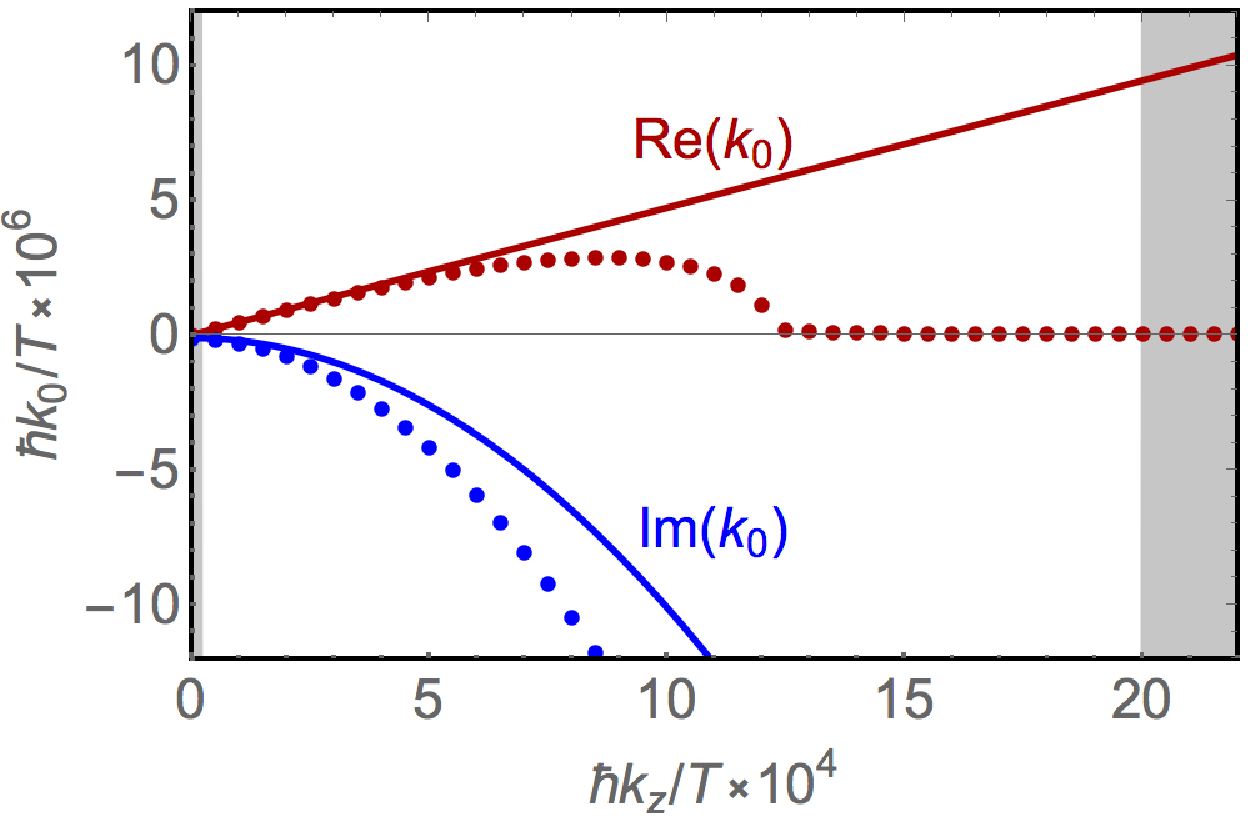}\hspace{0.01\textwidth}
    \includegraphics[width=0.32\textwidth]{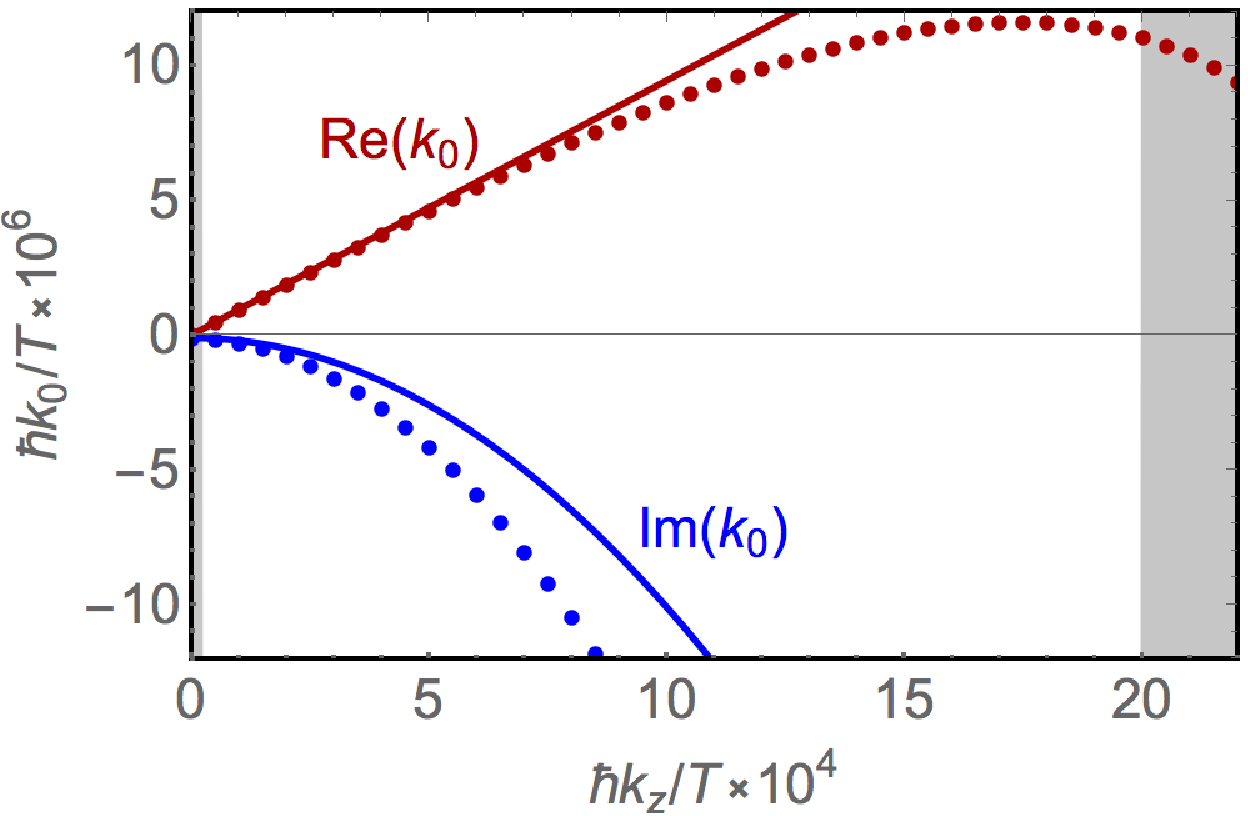}\hspace{0.01\textwidth}
    \includegraphics[width=0.32\textwidth]{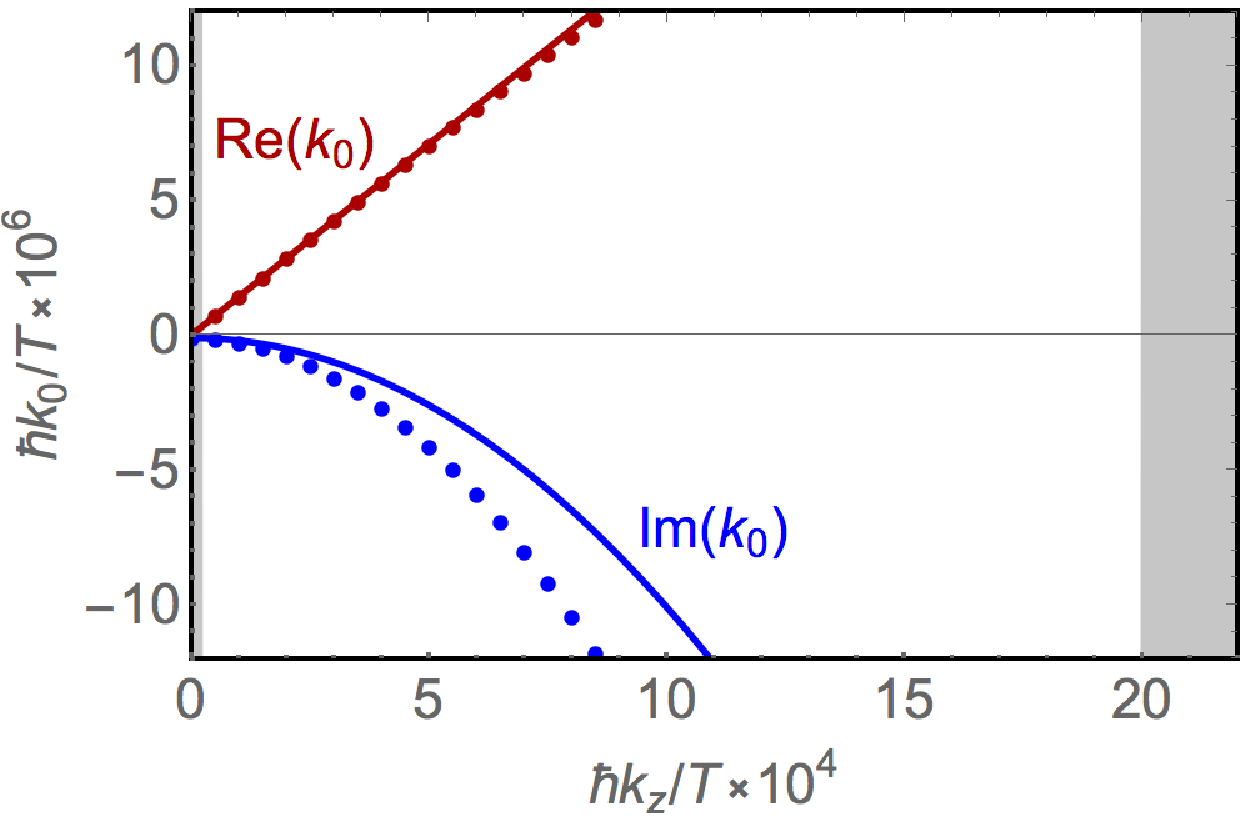}
    \caption{The comparison of the approximate analytical results (solid lines) for the real 
    (red lines and points) and imaginary (blue lines and points) parts of the energy for the 
    Alfv\'en waves with the corresponding numerical results (points) at $\Omega = 0$ for 
    three fixed values of the magnetic field: 
    $\hbar eB/T^2=0.5\times 10^{-3}$ (left panel), 
    $\hbar eB/T^2=10^{-3}$ (middle panel), and 
    $\hbar eB/T^2=1.5\times 10^{-3}$ (right panel). 
    The real and imaginary parts of the energy are shown in red and blue, respectively. 
    The other model parameters are $\tau T/\hbar=10^2$, $\mu/T = 10^{-4}$, and $\hbar k_\perp/T = 10^{-4}$.}
\label{fig:no_w_graph-1}
\end{figure}

Here it might be appropriate to note that the analytical expressions in Eq.~(\ref{eq:Alfven-wave-00}) were 
obtained by using the expansion in small parameter $\xi$, which was introduced in 
Eq.~(\ref{eq:scales_hot_plasma}) in order to separate different scales in the problem.
Therefore, while the corresponding dispersion relations provide good analytical approximations, they cannot 
be extended to the regions of very small and very large values of the wave vector. In order to support the 
validity of the approximate results obtained analytically, we compare them in Fig.~\ref{fig:no_w_graph-1}
with the dispersion relations found numerically. The panels in Fig.~\ref{fig:no_w_graph-1} 
show the results for  the real (red lines and points) and imaginary (blue lines and points) parts of the energy at 
three different fixed values of the magnetic field, i.e., $\hbar eB/T^2=0.5\times 10^{-3}$, 
$\hbar eB/T^2=1\times 10^{-3}$, and $\hbar eB/T^2=1.5\times 10^{-3}$, respectively. Note that, for the model 
parameters used, the hydrodynamic regime breaks down in the gray shaded regions at small and large values 
of $k_z$. As is clear from Fig.~\ref{fig:no_w_graph-1}, the analytical relations approximate well
the numerical results basically in the whole region where the real part of the energy remains larger than 
its imaginary part, i.e., $\mbox{Re}(k_0) \gtrsim \mbox{Im}(k_0)$.

In addition to the propagating modes, there are also two types of purely diffusive modes. 
The latter include the electric field decay mode with 
\begin{equation}
\label{eq:E-diffusive-00}
k_0 = - \frac{e^2}{9\hbar^3}i\tau T^2 
\end{equation}
and the chiral charge diffusive mode described by 
\begin{equation}
\label{eq:chiral-diffusive-00}
k_0 = - \frac{1}{3} i\tau k^2 - i\frac{27e^2B^2\hbar^2}{4\pi^4\tau T^4}. 
\end{equation}
It should be noted that there are three degenerate modes with the dispersion relation in 
Eq.~(\ref{eq:E-diffusive-00}), which correspond to three different polarizations of the electric 
field. The origin of these modes can be traced back to Ampere's law that takes a 
particularly simple approximate form $\partial_t \mathbf{E} +  \sigma_E\mathbf{E} \approx 0$.
This is also reconfirmed by the fact that the imaginary part in Eq.~(\ref{eq:E-diffusive-00}) is 
completely determined by the electrical conductivity, $\mbox{Im}(k_0)= - \sigma_E$. 

Before concluding this section, we would like to emphasize that there is no propagating 
mode in the spectrum that could be identified with the chiral magnetic wave. It is natural to ask, 
therefore, what is the reason for its absence. As we explain in Appendix~\ref{noCMW} in detail, the 
chiral magnetic wave is overdamped because of a high conductivity of hot plasma, which causes 
a rapid screening of the electric charge fluctuations and, thus, prevents the wave from forming.
While the situation is slightly more complicated in the strongly coupled quark-gluon 
plasma created in heavy-ion collisions, the chiral magnetic wave is still strongly overdamped 
due to the combined effects of electrical conductivity and charge diffusion \cite{Shovkovy:2018tks}.

\subsection{Charged plasma at $\Omega\neq 0$}

Let us now discuss the spectrum of hydrodynamic modes in the case of small, but nonzero vorticity.
As mentioned earlier, in order to simplify the analysis, we will limit ourselves only to linear order 
corrections to the dispersion relations in powers of $\Omega$. In such an approximation, the 
modes are classified by the same values of $k_\perp^{(i)}$ as at $\Omega=0$. Since there 
is no mixing of partial waves with different values of $k_\perp^{(i)}$, we will utilize a simpler notation 
$k_\perp$ instead of $k_\perp^{(i)}$ in the rest of the paper. 

Let us start by noting that the dispersion relation of the sound wave receives a linear correction 
in vorticity, i.e.,
	\begin{equation}
	k_0 = \frac{s_{e}k}{\sqrt{3}} - \frac{2}{15} i\tau k^2 + m\Omega\left( \frac{2}{3} 
	+ \frac{5e^2\mu^2}{14\pi^2\hbar^3 k^2} \right).
	\end{equation}
As in the case of vanishing vorticity, it remains a longitudinal wave. Its propagation is sustained 
primarily by oscillations of temperature $\delta T$ and velocity $\delta u^\mu$. However, at 
nonzero $\mu$, the wave could also excite small perturbations of the electromagnetic fields. 

To the leading linear order in the angular velocity $\Omega$, the dispersion relations of the 
Alfv\'en waves are given by the following approximate expression:
	\begin{equation}
	\label{eq:dispersion-Alfven}
		k_0^{(\pm)} = m\Omega + s_{e} k_z \sqrt{\frac{45\mathcal{B}_{\pm}^2\hbar^3}{7\pi^2T^4} 
		+ \left(\frac{\Omega}{k} - \frac{15e\mathcal{B}_{\pm}\mu}{14\pi^2T^2 k} \right)^2} 
		\pm k_z\left( \frac{\Omega}{k} - \frac{15e\mathcal{B}_{\pm}\mu}{14\pi^2T^2 k} \right) 
		- i\tau k^2\left(\frac{1}{10} + \frac{9\hbar^3}{2e^2\tau^2T^2}\right),
	\end{equation}
where we introduced the following shorthand notation:
\begin{equation} 
\label{eq:modified_magnetic_field2}
	\mathcal B_{\pm} = B - \frac{en_{\rm eq}\Omega}{6k_\perp^2} \left[2(m\pm 1)(m\pm2)+k_\perp^2R^2\right] .
\end{equation}
Note that we used $n_{\rm bg}=n_{\rm eq}$, which enforces the electric charge neutrality in the plasma at 
$\Omega =0$, see Eq.~(\ref{eq:n_vs_r_perp}). As we can see, there are four different branches of Alfv\'en 
waves. They are determined by two different circular polarizations, labeled by the 
$\pm$ signs in Eq.~(\ref{eq:dispersion-Alfven}), and two directions of propagation with respect to the $z$ 
axis. The latter are formally distinguished by $s_{e}=\pm 1$. (Recall that both the background magnetic field 
and the axis of rotation are parallel to the $z$ axis.) 

\begin{figure}[t]
    \includegraphics[width=0.32\textwidth]{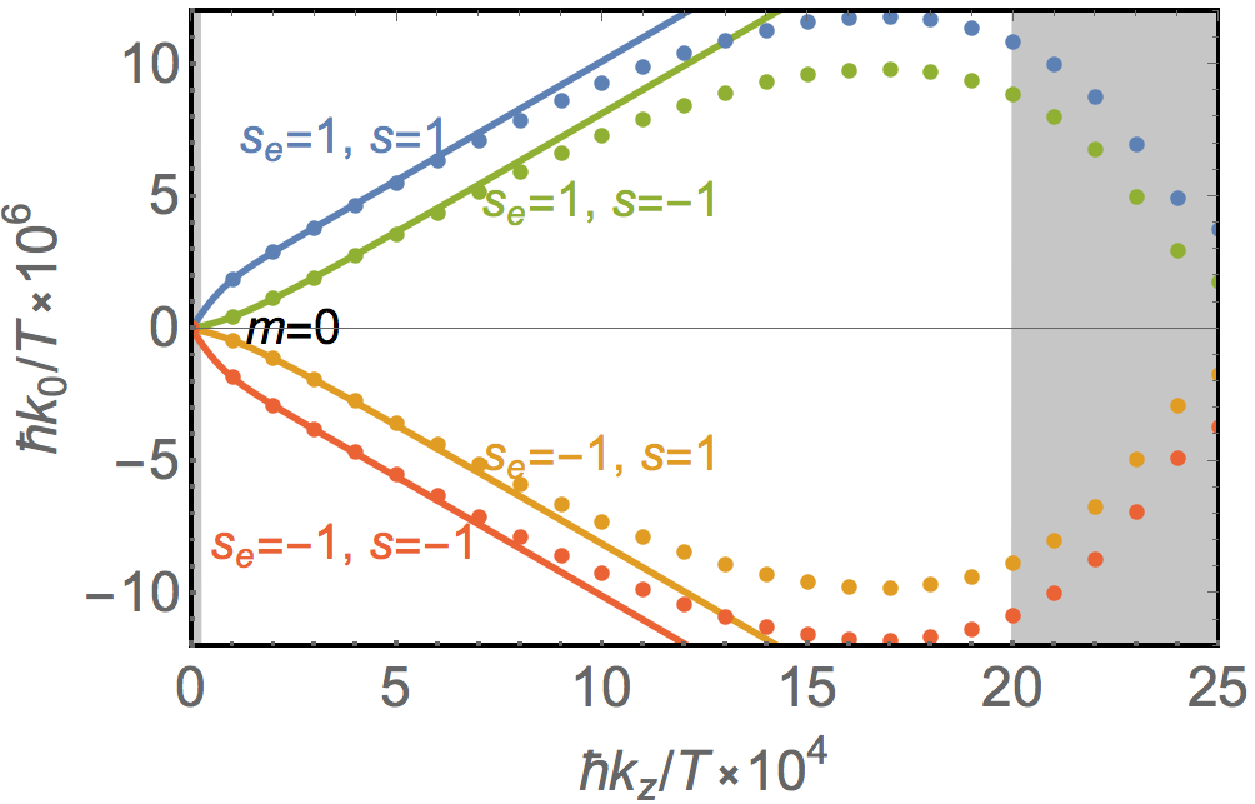}\hspace{0.01\textwidth}
    \includegraphics[width=0.32\textwidth]{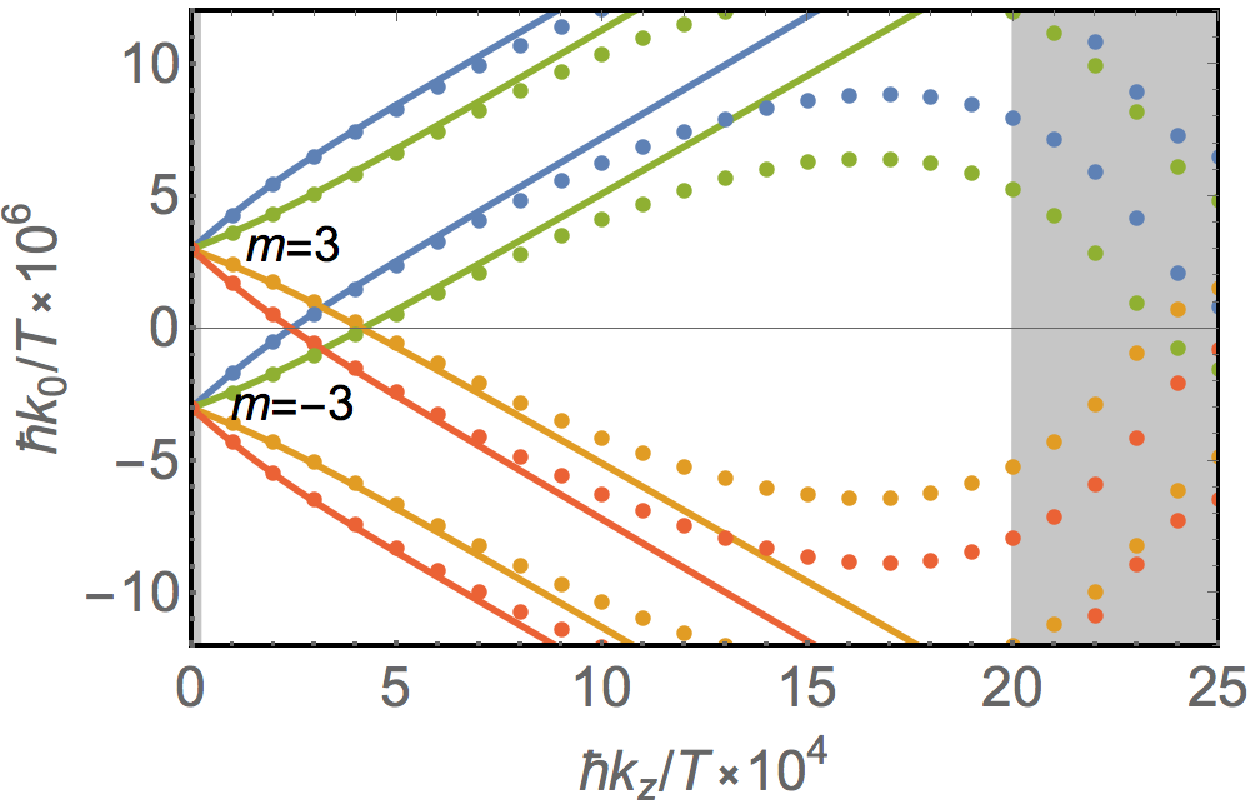}\hspace{0.01\textwidth}
    \includegraphics[width=0.32\textwidth]{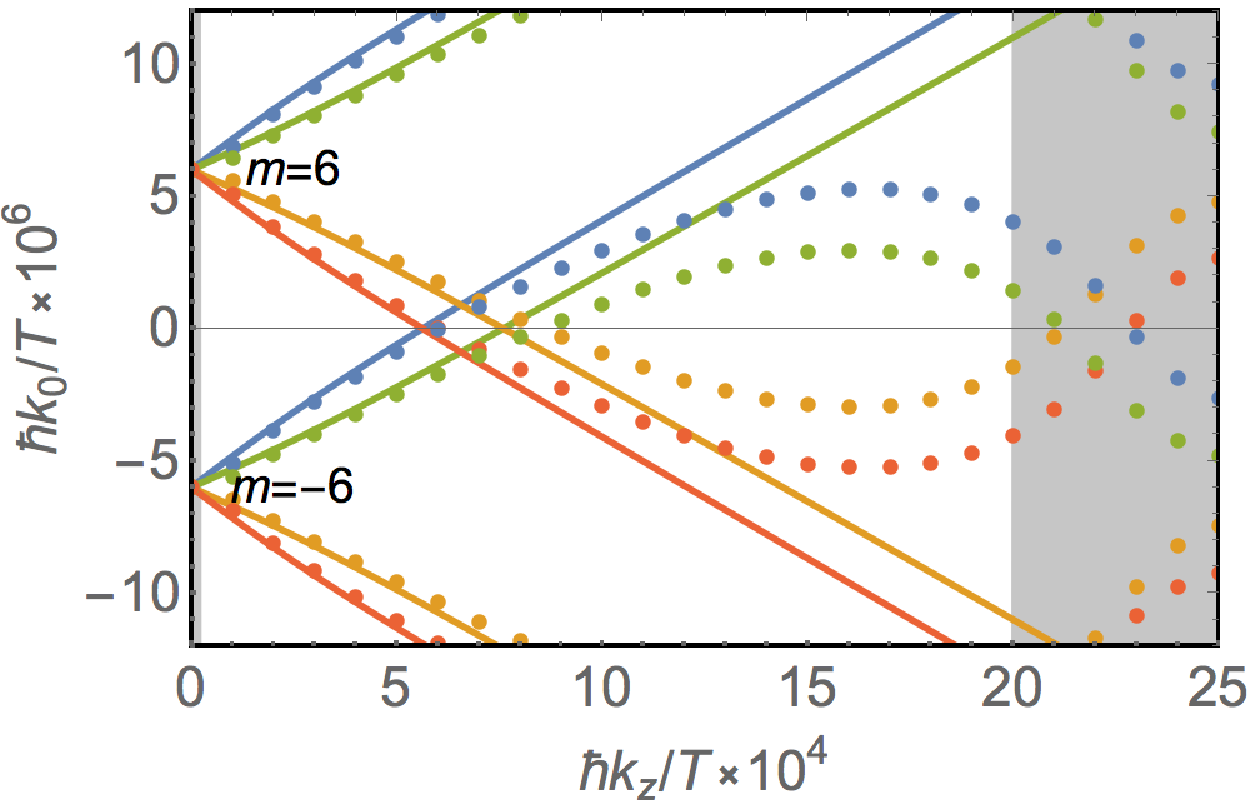}
    \caption{The real parts of the energies of the Alfv\'en waves with different values of the angular 
    momentum: $m=0$ (left panel), $m=\pm3$ (middle panel), and $m=\pm6$ (right panel). The other 
    model parameters are $\tau T/\hbar=10^2$, $\mu/ T=10^{-4}$, $\hbar\Omega/T= 10^{-6}$, 
    $\hbar eB/T^2=10^{-3}$, $RT/\hbar=10^{4}\alpha_{0,1}\approx 2.4\times10^{4}$, and 
    $k_\perp = \alpha_{m,1}/R$.}
	\label{fig:Alfven_waves_omega}
\end{figure}

By comparing the result in Eq.~(\ref{eq:dispersion-Alfven}) with the dispersion relation at $\Omega=0$, 
given by Eq.~(\ref{eq:Alfven-wave-00}), we see that the inclusion of vorticity lifts the degeneracy of 
modes propagating in opposite directions with respect to the background magnetic field (and/or vorticity). 
Moreover, we also find that the propagation of these types of waves is modified by the chiral vortical effect. 
This is most pronounced in the case of small $k_z$ (i.e., $k_z \lesssim k_\perp$). In such a case, therefore,
it might be suitable to call these modes the Alfv\'en-vortical waves. The dispersion relations for several 
modes with different values of the angular momentum $m$ are shown in Fig.~\ref{fig:Alfven_waves_omega}. 
(The hydrodynamic regime breaks down in the gray shaded regions at small and large values of $k_z$.)
Note that the approximate analytical expressions (represented by solid lines) are in good agreement 
with the numerical results (shown with dots) in the region of small momenta.

It is interesting to note that the energies of the circularly polarized Alfv\'en waves depend on the magnetic 
field only via the combinations $\mathcal{B}_{\pm}$, defined in Eq.~(\ref{eq:modified_magnetic_field2}). 
This means that one of the circularly polarized waves with a fixed angular momentum $m$ could be 
fine-tuned (e.g., by adjusting the magnetic field so that $\mathcal{B}_{-}=0$) into a pure chiral vortical wave 
with the dispersion relation given by $k_0 \approx m\Omega + (s_{e}-1)\frac{k_z}{k}\Omega - \frac{1}{10}i\tau k^2$. 

For completeness, let us now briefly discuss the diffusive modes. The electric field decay mode remains 
the same as in Eq.~(\ref{eq:E-diffusive-00}). As for the chiral charge diffusive mode, at nonzero $\Omega$, 
it is given by
\begin{equation}
	\label{eq:diffusive-chiral}
		k_0 = m\Omega - i\frac{27(e\mathcal{B}_0)^2\hbar^2}{4\pi^4\tau T^4} - \frac{1}{3} i\tau k^2,
\end{equation}
where we introduced the following shorthand notation:
\begin{equation}
		\label{eq:modified_magnetic_field1}
		\mathcal B_0 = B - \frac{en_{\rm eq}\Omega}{6k_\perp^2} \left[2(m-1)(m+1)+k_\perp^2R^2\right].
\end{equation}
Note that the energy of the chiral charge diffusive mode has a nonzero real part proportional to 
$\Omega$. One might speculate, therefore, that under certain conditions and beyond the leading 
order in $\Omega$, it may even become a propagating mode. As is easy to check, the chiral charge 
diffusive mode is determined primarily by the corresponding continuity relation. At nonzero $\Omega$, 
in particular, the latter can be approximately given by 
\begin{equation}
		\partial_0 n_5 = -\frac{\tau}{3}\nabla^2 n_5 - (\bm{\Omega}\times\bm{\nabla} n_5)\cdot \bm{r}_\perp .
\end{equation}
By solving this, we can indeed reproduce the first and the last terms in the dispersion relation 
(\ref{eq:diffusive-chiral}). This reconfirms that, while other degrees of freedom may influence this diffusive 
mode in principle, their role is minor.

\subsection{Charged plasma at $\Omega\neq 0$ without dynamical electromagnetic fields}

As we argued in Sec.~\ref{sec:theoretical_framework} by using rather general arguments,
a self-consistent treatment of charged chiral plasma requires a proper inclusion of fully dynamical 
electromagnetic fields. In this subsection, we test the validity of such a claim by performing a 
comparative study without the inclusion of the dynamical fields. In order to achieve such a regime 
in the charged chiral plasma, we will assume that the matter is affected only by the static
{\em background} magnetic field. No additional background electric fields can be allowed 
because those would drive dissipative Ohm currents. When neglecting dynamically induced 
electromagnetic fields, there will be no effect from such fields on the plasma modes. As we will 
see, many signature features of hydrodynamic modes will be lost in such an approximation. 
This finding, therefore, will reconfirm the importance of accounting for the dynamical fields.

When dynamical electromagnetic fields are neglected, the Maxwell equations play no role in 
determining the properties of hydrodynamic modes. This means that one is left with the system 
of only six continuity equations (\ref{eq:continuity-1})--(\ref{eq:continuity-3}). To leading order 
in $\Omega$, we can assume that the background values of the chemical potential and 
temperature are spatially uniform. In the absence of the background electric field, the chiral
anomaly is effectively switched off. This means that both fermion number (electric) and chiral currents are 
conserved and affect collective modes in similar ways. 

We find that there are three kinds of propagating hydrodynamic modes: a longitudinal sound
wave, a transverse vortical wave, and a transverse chiral magnetic wave. (Note 
that the terms longitudinal and transverse refer to the direction of fluid flow oscillations with 
respect to the wave vector.) Their dispersion 
relations read
	\begin{eqnarray}
	\label{eq:s-mode-noEM}
			k_0 &=& \frac{s_{e}k}{\sqrt{3}}  + \frac{2}{3}m\Omega - \frac{2}{15} i\tau k^2, \\
	\label{eq:v-mode-noEM}
			k_0 &=& m\Omega + s_{e}\frac{2k_z\Omega}{k} - \frac{1}{5}i k^2\tau, \\
	\label{eq:mv-mode-noEM}
			k_0 &=& m\Omega + s_{e}\frac{3e\mathcal{B}_0\hbar k_z}{2\pi^2T^2} - \frac{1}{3}i k^2\tau,
	\end{eqnarray}
respectively. As expected, the sound mode is not affected much by omitting dynamical electromagnetic 
fields. However, it did become completely independent of the chemical potential. This should have been 
expected though since oscillations of local electric fields from charge density perturbations are artificially 
switched off now. 

The modes in Eq.~(\ref{eq:v-mode-noEM}) are substitutes for the Alfv\'en waves (\ref{eq:dispersion-Alfven}) 
in the fully dynamical case. 
By comparing their dispersion relations, we clearly see that the modes are drastically different. This is 
further confirmed by reviewing the underlying nature of the two sets of modes. For example, the  
vortical wave (\ref{eq:v-mode-noEM}) is driven almost exclusively by velocity perturbations $\delta u^\mu$. 
A pair of the weakly damped chiral magnetic waves (\ref{eq:mv-mode-noEM}) is driven by 
oscillations of either left-handed ($s_{e}=-1$) or right-handed ($s_{e}=1$) particles. As expected, they 
replace a pair of diffusive modes found in the dynamical regime in the previous subsection. At 
$\mathcal{B}_0 = 0$, they turn again into diffusive electric and chiral charge waves, driven by the 
perturbations of $\delta\mu$ and $\delta\mu_5$, respectively.

To summarize the results of this subsection, we find that neglecting dynamical electromagnetic fields 
has a profound effect on the spectrum of collective modes in the hydrodynamic regime. One of them 
is the appearance of propagating chiral magnetic waves, which are absent in the charged
chiral plasma with dynamical electromagnetism. The other qualitative difference is the absence of the
correct Alfv\'en waves, which are replaced by the vortical wave with a rather different dispersion relation.

\subsection{Plasma of neutral particles at $\Omega\neq 0$}

For completeness, it might also be interesting to discuss the hydrodynamic modes in a chiral plasma made 
of neutral particles. Clearly, the Maxwell equations play no role in this case. The hydrodynamic description 
is governed by Eqs.~(\ref{eq:continuity-1})--(\ref{eq:continuity-3}) with the vanishing electromagnetic 
fields. To leading order in $\Omega$, the values of the chemical potential $\mu$ and temperature $T$ for 
the uniformly rotating neutral plasma can be assumed constant in the global equilibrium. Moreover, we can
even include an arbitrary nonzero axial chemical potential $\mu_5$. 

Among the propagating modes in neutral plasma, we find the usual longitudinal sound wave with the 
dispersion relation given by
	\begin{equation}
		k_0 = \frac{s_{e}k}{\sqrt{3}}  + \frac{2}{3}m\Omega - \frac{2}{15} i\tau k^2 ,
	\end{equation}
where $s_{e}=\pm 1$, and a single circularly polarized vortical wave with the dispersion relation 
	\begin{equation}
	\begin{aligned}
		k_0 = m\Omega + s_{e}\frac{2k_z\Omega}{k} - \frac{1}{5}i\tau k^2.
	\end{aligned}
	\end{equation}
It should be emphasized that the sign $s_{e}=\pm 1$ defines one of the two possible directions 
of propagation of the vortical wave with respect to the $z$ axis. For each direction of the 
propagation, though, there is only one circularly polarized mode. This differs qualitatively 
from the Alfv\'en-vortical waves in charged plasmas which have two propagating circularly 
polarized modes for each direction. 

The dispersions of diffusive modes associated with the fermion number charge and the 
chiral charge are degenerate and resemble the zero-field limit of that in Eq.~(\ref{eq:diffusive-chiral}). 
In particular, they are given by
	\begin{equation}
		k_0 = m\Omega - \frac{1}{3}i\tau k^2.
	\end{equation}
It is rather interesting to point out that, to leading order in $\Omega$, the equilibrium values 
of the chemical potentials $\mu$ and $\mu_5$ do not affect the energy spectrum of the 
modes in the neutral chiral plasma. The spectra also are not dependent on temperature.

\section{Hydrodynamic modes in dense plasma}
\label{sec:Modes-highMu}

In this section, we study the spectrum of hydrodynamic modes in a chiral rotating plasma at high 
density, i.e., $\mu\gg T$. Such a regime could be realized in compact stars and, perhaps, also in 
Dirac and Weyl materials, in which low-energy electron quasiparticles behave as pseudorelativistic 
chiral fermions. 

As in the case of hot plasma in the previous section, it is convenient to use a specific hierarchy 
of all relevant scales in the problem by relating them via a single small parameter $\xi \simeq 10^{-2}$.
In the case of dense plasma, we will use 
\begin{equation}
\label{eq:scales_dense_plasma}
	l_{\rm mfp} \Omega  \simeq \xi^{5/2} , 
	\qquad
	\xi^{3/2}\simeq\frac{l_{\rm mfp}}{R}\lesssim k l_{\rm mfp} \lesssim\xi^{1/2},
	\qquad
	\frac{l_{\rm mfp}}{l_B} \simeq \xi^{-1/4} ,
	\qquad
	\frac{l_{\rm mfp}}{l_{d}} \simeq \xi^{-1} .
\end{equation}
It should be noted that here we consider even smaller vorticity in relative terms than in hot chiral plasma, 
see Eq.~(\ref{eq:scales_hot_plasma}). This is motivated by the need to keep the vorticity corrections to 
the magnetic field~(\ref{eq:modified_magnetic_field1}) and (\ref{eq:modified_magnetic_field2}) under 
control in the regime of a large fermion number density. While considering the high-density 
regime, it is instructive to see how the effects of a small temperature start showing up in the spectrum 
of collective modes. For this purpose, we include a nonzero temperature, but assume that its value is  
small, e.g., $T \simeq \xi \mu$ (or, equivalently, $T l_{d}\simeq \xi\hbar$, where 
$l_{d}= \hbar/\mu$).

\subsection{Charged plasma at $\Omega= 0$}

Before considering the case of nonzero vorticity, we review the spectrum of hydrodynamic modes
in the case $\Omega=0$ by solving the linearized system of equations 
(\ref{eq:system-simple-start})--(\ref{eq:system-simple-end}). As expected, we find that there are two 
kinds of propagating modes at $\Omega=0$, namely, plasmons and helicons. Also, there are two 
diffusive modes with identical dispersions, $k_0 = - i\tau k^2/3$. Note that, in the regime of dense 
matter, there are no usual sound waves. They are replaced by plasmons. Similarly, the Alfv\'en waves
are morphed into helicons. 

Plasmons describe the propagation of charge oscillations sustained by dynamically induced 
electric fields. As is well known, their frequency for the ultrarelativistic plasma without background 
fields and rotation is given by $\omega_\textrm{PL} = \frac{e\mu}{\sqrt{3}\pi\hbar^{3/2}}$.
The plasmon can have three degenerate modes with different polarizations. 
We find that the degeneracy is lifted by the magnetic field. Indeed, from our 
linearized system of equations, we find that the dispersion relations of the plasmon modes are 
given by
\begin{equation}
\label{eq:plasmon-00}
k_0^{(s)} = s_{e} \frac{e\mu}{\sqrt{3}\pi\hbar^{3/2}} + s\frac{e B }{2\mu} 
- \frac{ie^2\tau T^2}{18\hbar^3} - \frac{1}{10}i\tau k^2, 
\end{equation}
where $s_{e}=\pm1$ and $s=-1,0,1$. In terms of the hydrodynamic variables, plasmons are primarily driven by 
the oscillations of flow velocity $\delta u_{\pm}$ and electric field $\delta E_{\pm}$. For $s=\pm 1$, the 
modes have clockwise (with nonzero $\delta u_+$ and $\delta E_+$) and anticlockwise 
(with nonzero $\delta u_-$ and $\delta E_-$) circular polarizations, respectively. The case
of $s=0$ corresponds to the mode with the linear polarization in the $z$ direction. 

The dispersion relations of the helicon mode are given by 
\begin{equation}
\label{eq:helicon-00}
k_0 = s_{e}\frac{3\pi^2eBkk_z\hbar^3}{e^2 \mu^3} - \frac{3\pi^2\hbar^3}{5e^2 \mu^2}i\tau k^4,
\end{equation}
where $s_{e}=\pm 1$. One of the signature features of such magnetohydrodynamic modes is their 
quadratic dispersion relations. They are circularly polarized with a given handedness, determined 
by the sign of $s_e$. In terms of the hydrodynamic variables, the propagation of helicons is driven 
primarily by oscillations of flow velocity $\delta u_{\pm}$, temperature $\delta T$, and magnetic 
field $\delta B_{\pm}$. It might be appropriate to mention that, for a typical choice of the model 
parameters with the hierarchy of scales in Eq.~(\ref{eq:scales_dense_plasma}), the helicons 
are well-defined propagating (rather than overdamped) modes in the whole region of momenta, 
$\xi^{3/2}\lesssim k l_{\rm mfp}\lesssim\xi^{1/2}$. This remains also marginally true even 
for a range of somewhat weaker magnetic fields, provided $l_{\rm mfp}/l_B \gtrsim 1$. However, 
in the case of very weak fields, the helicons will become overdamped already at some intermediate 
values of the wave vector (e.g., $k l_{\rm mfp} \simeq \xi^{3/4}$ when $l_{\rm mfp}/l_B \simeq \xi^{1/4}$).

\subsection{Charged plasma at $\Omega\neq 0$}

Let us now proceed to the case of a rotating chiral plasma. As in the case of hot plasma in the previous
section, we will study the modifications of hydrodynamic modes up to linear order in $\Omega$. In this
case, the modes are classified by well-defined transverse wave vectors $k_\perp$. (Recall that the 
corresponding values are discretized $k_\perp^{(i)} =\alpha_{m,i}/R$, but we will omit the superscript
in order to simplify the presentation.)

\begin{figure}[t]
		\includegraphics[width=0.5\textwidth]{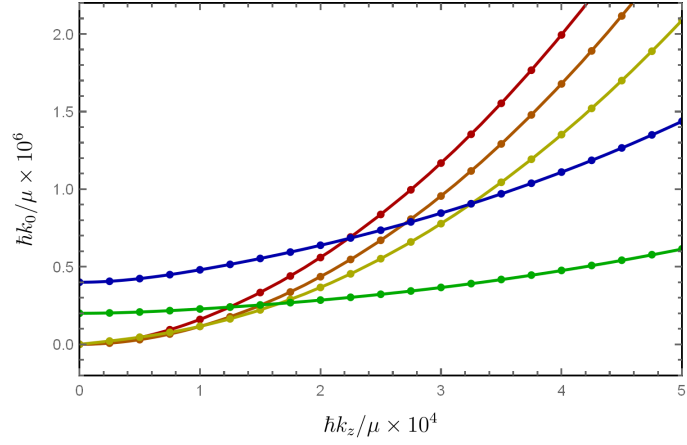}
		\caption{The positive branches of the real part of helicon energies for several values of 
		the angular momentum, i.e., 
		$m=-4$ (red line), $m=-2$ (orange line), $m=0$ (olive line), 
		$m=2$ (green line), and $m=4$ (blue line). The other model parameters are 
		$\hbar eB/\mu^2 \approx 2\times 10^{-3}$, $\tau \mu/\hbar=100$, 
		$ \hbar\Omega/\mu= 10^{-7}$, and $k_\perp = \alpha_{m,1}/R$.}
	\label{fig:high_density_helicon-1}
	\end{figure}

We start by noting that the $\Omega =0$ plasmon dispersion relation given by Eq.~(\ref{eq:plasmon-00}) 
remains almost the same, but the magnetic field $B$ in the subleading term is replaced by $\mathcal{B}_s$, 
i.e., 
\begin{equation}
\label{eq:plasmon-omega}
k_0^{(s)} = s_{e} \frac{e\mu}{\sqrt{3}\pi\hbar^{3/2}} + s\frac{e\mathcal{B}_s}{2\mu} 
- \frac{ie^2\tau T^2}{18\hbar^3} - \frac{1}{10}i\tau k^2 . 
\end{equation}
As for the helicon mode, its dispersion relation becomes
	\begin{equation}
	\label{eq:high-density-helicon}
		k_0 = m\Omega\left( \frac{1}{2} - \frac{k_z^2}{k_\perp^2} \right) + s_{e}\sqrt{\frac{m^2\Omega^2}{4} + \frac{9\pi^4(\mathcal{B}_{+} + \mathcal{B}_{-})^2k^2k_z^2\hbar^5}{4\mu^6}} 
		- \frac{3\pi^2\hbar^3}{5e^2\mu^2}i\tau k^4,
	\end{equation}
where $s_{e}=\pm 1$. As is easy to see, the positive branches of the real part of energy (i.e., $s_{e}=+1$) 
are gapped for $m>0$ and gapless for $m\le 0$. Concerning the case of $m>0$, the values of the 
gaps are determined by the angular velocity, $m\Omega$. A typical spectrum is illustrated in
Fig.~\ref{fig:high_density_helicon-1}, where the dispersion relations for several fixed values of 
the angular momentum ($m=-4,-2,0,2,4$) are shown. Note that, in the figure, we zoomed into 
the region of very small energies. By taking into account that $\Omega$ is very small, it should be
clear that the complete spectrum contains a nearly continuous range of gaps. 
	
It is interesting to note that for certain values of the angular momentum (when the magnetic field is fixed),
the effective field $(\mathcal{B}_{+} + \mathcal{B}_{-})$ could become very small, or even zero. In this
case, the first term in the dispersion relation (\ref{eq:high-density-helicon}) dominates and leads to a 
quadratic dependence on $k_z$ with a negative overall coefficient. (Note that the energies for the 
branches with negative values of $m$ have opposite signs.) 

\begin{figure}[b]
	\includegraphics[width=0.5\textwidth]{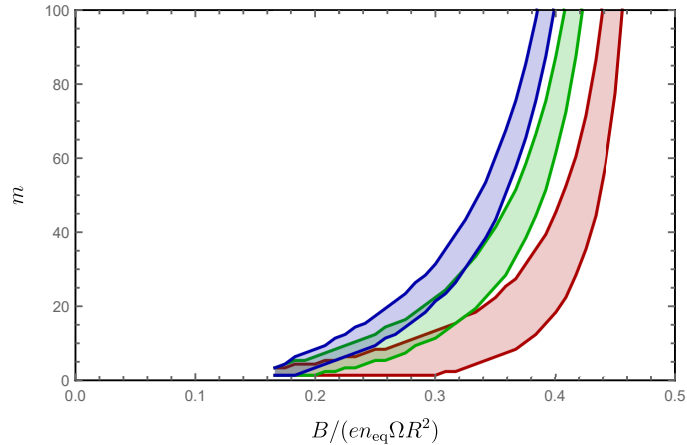}
	\caption{The ranges of angular momenta $m$ for which helicon modes can have 
	negative group velocity. The colored bands correspond to three smallest values of 
	the transverse momenta $k_\perp^{(i)}$ with $i=1,2,3$ (from red to blue, respectively).}
	\label{fig:high_density_Alfven-m-bands}
\end{figure}

From a physics viewpoint, the negative curvature of the dispersion relation as a function of 
$k_z$ implies that the group velocity of such modes $v_z$ is negative. This is a rather interesting 
feature that, potentially, could be important in applications. By analyzing the analytical expression 
in Eq.~(\ref{eq:high-density-helicon}), we find, however, that a negative group velocity can be 
realized only for rather large magnetic fields. Indeed, by making use of the properties of the Bessel 
functions, we find that the negative slope is possible only when the magnetic field lies in the 
following window: $en_{\rm eq}R^2\Omega/6<B<en_{\rm eq}R^2\Omega/2$. This corresponds 
to the dimensionless ratio $l_{\rm mfp}/l_{B} \simeq \xi^{-7/4}(e/\sqrt{\hbar})$, which is 
substantially larger than $1$ even for a rather small coupling constant, $e/\sqrt{\hbar} 
= 1/\sqrt{137}$. The ranges of angular momenta $m$ of the modes with $v_z<0$ are 
illustrated graphically in Fig.~\ref{fig:high_density_Alfven-m-bands}. There we show three
colored bands that correspond to the three smallest values of the transverse momenta 
$k_\perp^{(i)}$ with $i=1,2,3$. As we see, the ranges of bands in $m$ rise very steeply 
as $B/(en_{\rm eq}R^2\Omega)$ approaches $1/2$. The bands also have a tendency 
to shift upwards with increasing $k_\perp^{(i)}$.

In addition to the propagating modes, there is also a pair of overdamped modes associated with the 
diffusion of chiral charge and energy, i.e., 
	\begin{eqnarray}
	\label{eq:diffusion-mode-dense-1}
		k_0 &=& m\Omega - \frac{1}{3}i\tau k^2 ,\\
	\label{eq:diffusion-mode-dense-2}
		k_0 &=& m\Omega\left(1-\frac{e^2\mu^2\tau^2}{9\pi^2\hbar^3}\right) - \frac{1}{3}i\tau k^2,
	\end{eqnarray}
respectively. In the hydrodynamic regime defined by the hierarchy of length scales in 
Eq.~(\ref{eq:scales_dense_plasma}), neither of these modes has a chance of becoming 
a well-resolved propagating mode.

\subsection{Charged plasma at $\Omega\neq 0$ without dynamical electromagnetic fields}

As in the case of hot plasma in the previous section, here it is also instructive to verify that 
the description of hydrodynamic modes is substantially modified in the background-field 
approximation, i.e., when the dynamical electromagnetic fields are neglected. 

Switching off the dynamical electromagnetic fields is equivalent to ignoring the Maxwell 
equations. Then the simplified system of the six linearized equations contains only continuity 
equations (\ref{eq:continuity-1})--(\ref{eq:continuity-3}). Here we will concentrate on the 
propagating modes, which are particularly interesting. As for the diffusive modes, one can 
verify that there are two degenerate modes with the dispersion relation given by 
Eq.~(\ref{eq:diffusion-mode-dense-1}). 

One of the immediate consequences of the approximation without dynamical electromagnetic 
fields is the absence of the plasmons in the spectrum. They are replaced by the sound waves
with the energy given by
	\begin{equation}
		k_0 =  \frac{s_{e}k}{\sqrt{3}} + \frac{2}{3}m\Omega - \frac{2}{15} i\tau k^2.
	\end{equation}
The underlying physics of such a dramatic change is clear. While neglecting the Gauss law, 
local oscillations of the electric charge density do not produce any electric field, 
resulting in a gapless sound wave exactly as in the case of plasma made of neutral particles. 

While the helicon remains in the spectrum, its dispersion relation is substantially modified. 
In particular, its energy in the background-field approximation is given by 
	\begin{equation}
	k_0 = m\Omega + \frac{s_{e}}{5\mu}e\mathcal{B}_skk_z\tau^2 - \frac{1}{5}i\tau k^2,
	\end{equation}
where $s_{e}=\pm1$. In essence, this is a purely hydrodynamic mode, which is driven by 
oscillations of the fluid velocity. Its propagation is accompanied by oscillations of temperature, 
as well as small oscillations of the electric and chiral chemical potentials.

\subsection{Chiral plasma of neutral particles at $\Omega\neq 0$}

For completeness, let us also discuss the case of chiral plasma made of neutral particles.
Since the chiral anomaly is absent in this case, it is straightforward to include a nonzero 
chiral chemical potential $\mu_5$ along with the fermion number chemical potential $\mu$. 
Note, however, that none of the hydrodynamic modes in this regime will be modified
by the values of $\mu_5$ and $\mu$. This situation is qualitatively the same as in the case of 
hot plasma made of neutral particles. Also, as in that case, the spectra do not depend on temperature.

The energies of the sound and vortical waves are given by the following expressions:
	\begin{eqnarray}
		k_0 &=& \frac{s_{e}k}{\sqrt{3}}  + \frac{2}{3}m\Omega - \frac{2}{15} i\tau k^2,\\
		k_0 &=& m\Omega + s_{e}\frac{2k_z\Omega}{k} - \frac{1}{5}i\tau k^2,
	\end{eqnarray}
respectively. While the former is a longitudinal wave, the latter is a transverse 
circularly polarized one. There is also a pair of degenerate diffusion modes with 
the dispersion relation given by Eq.~(\ref{eq:diffusion-mode-dense-1}).

\section{Summary}
\label{sec:Summary}

In this paper, we used a covariant formulation of the chiral kinetic theory \cite{Hidaka:2016yjf,Hidaka:2017auj} 
in the relaxation-time approximation in order to derive the first-order dissipative hydrodynamics equations 
for a magnetized chiral plasma with nonzero vorticity. By noting that dynamical electromagnetism should 
play a profound role in such a relativistic charged plasma, we argued that the complete set of hydrodynamic
equations should necessarily include the appropriate Maxwell equations. 

Furthermore, by utilizing the corresponding hydrodynamic framework with dynamical electromagnetism,
we derived the complete spectrum of hydrodynamic modes almost completely by analytical methods. The 
task of solving the linearized equations was greatly simplified by assuming that the plasma is confined 
to a cylindrical region of finite radius $R$ rotating uniformly with a small angular velocity $\Omega$. 
Then, by imposing suitable boundary conditions, at linear order in $\Omega$, we were able to classify 
all modes in terms of the angular momenta $m$ and the transverse wave vectors 
$k_\perp^{(i)}$. The latter, in particular, are determined by the inverse radius $R^{-1}$, multiplied  
by the zeros of the Bessel functions. Such a setup provides a rigorous and 
systematic way not only for including the combined effects of the background magnetic field and rotation
but also for treating systems of finite size.

One of the critical details in our analysis of hydrodynamic modes is self-consistent determination 
of the unperturbed (equilibrium) state for a charged plasma rotating with a constant angular velocity. 
As we showed, relativistic effects force the corresponding state to be nonuniform in the radial direction. 
The situation is further complicated by the possible presence of a static background of nonchiral particles, 
in which case the nonuniformity shows up already at the linear order in $\Omega$. We verified that 
(i) the electric and magnetic forces on any element of the rotating plasma are exactly compensated, 
and (ii) there are no dissipative hydrodynamic processes in the correct equilibrium state.

The main results of this paper also include detailed spectra of hydrodynamic modes in the 
regimes of both high temperature and high density. In the case of hot plasmas, the main propagating modes 
are the sound and Alfv\'en waves. At high density, on the other hand, the corresponding modes are 
the plasmons and the helicons. In all regimes, the dispersion relations are affected in a rather nontrivial 
way by a nonzero angular velocity $\Omega$. One of the notable exceptions is the plasmon, which 
remains intact up to the linear order in $\Omega$. It is almost certain, of course, that its energy will 
be changed at higher orders in $\Omega$, but our approximation 
was insufficient to reconfirm this explicitly. We also tested directly that dynamical electromagnetism 
plays an important role in determining qualitative properties of all well-resolved (i.e., not 
overdamped) propagating modes. This was done by comparing the results with and without the
inclusion of the dynamical fields.  We found that the spectra of collective modes also include a 
number of diffusive (overdamped) modes. 

At the end, we would like to argue that the results of this study could be very important in guiding the 
Sisyphean task of searching for possible chiral anomalous effects in various forms of relativistic plasmas, 
ranging from the hot plasma in the early universe to cold matter in compact stars. Of particular interest 
are the elusive signatures of the chiral magnetic and vortical effects in the quark-gluon plasma 
produced by ultrarelativistic heavy-ion collisions. Generically, the corresponding signatures are expected 
to be seen in various multiparticle correlators  \cite{Liao:2014ava,Kharzeev:2015znc,Huang:2015oca}. 
Often, however, the predictions for such observables are based on simplified descriptions of anomalous 
collective modes in chiral plasma in the background-field approximation. This is the case, for 
example, in relation to predictions for the quadrupole charged-particle correlations associated with the 
chiral magnetic wave \cite{Gorbar:2011ya,Burnier:2011bf}. As we argued in this study, the background-field 
approximation is not reliable. In fact, our results clearly demonstrate that there are no chiral 
vortical and chiral magnetic waves in the spectrum of collective excitations. They either become 
diffusive or morph into the conventional plasma modes such as the sound waves, the Alfv\'en 
waves, the helicons, and the plasmons. 

In connection to the chiral magnetic wave, in particular, we found that such a mode becomes 
overdamped almost in all realistic regimes of hot and dense plasmas after the effects of dynamical 
electromagnetism are taken into account. From a physics viewpoint, the propagation of the wave 
is badly disrupted by the high electrical conductivity $\sigma_E$ that causes a rapid screening 
of the electric charge fluctuations. In the case of hot plasma, for example, the corresponding 
screening timescale is of order $1/\sigma_E\sim e^2 /T$, which is much shorter than the timescale 
$T^2/(eB_0 k )$ needed for the chiral magnetic/separation effects to initiate the wave. (Alternatively,
the screening rate $\sigma_E\sim T/e^2$ is much higher than the frequency $k eB_0/T^2$ of the 
presumed chiral magnetic wave.)

In application to heavy-ion physics, this means that there are no
theoretical foundations to expect that chiral magnetic waves are possible in hot quark-gluon
plasma \cite{Shovkovy:2018tks}. Consequently, we conclude that the observations of the charge-dependent flows or, 
in other words, quadrupole charged-particle correlations \cite{Ke:2012qb,Adamczyk:2013kcb,
Adamczyk:2015eqo,Adam:2015vje,Sirunyan:2017tax} are unlikely to be connected 
with the chiral magnetic wave, or any anomalous physics for that matter. Moreover, this might 
explain why the experimental effort to extract the signal from the background appears to be 
so difficult \cite{Adam:2015vje,Sirunyan:2017tax}.

In the end, let us mention that the collective modes studied here may 
be of relevance also in the early universe. Generically, of course, vector modes and vorticity 
are absent in inflationary models, remain small at all cosmological epochs, and should decay 
almost completely during the time of matter domination (for a review see, e.g., Ref.~\cite{Gorbunov-Rubakov}). 
Nevertheless, by noting that the primordial magnetic fields might be connected to vortical fluid 
perturbations, the hydrodynamic modes and vorticity could also be relevant for certain aspects 
of physics in the early universe. The study of the corresponding effects is beyond the scope of 
this paper, however.

\begin{acknowledgments}
The authors thank Naoki Yamamoto for valuable comments regarding the earlier version 
of the manuscript. I.A.S. also thanks Dmitri Kharzeev, Nathan Kleeorin, Igor Rogachevskii, 
and Pavlo Sukhachov for numerous discussions related to the topic of this work during the Nordita scientific 
program ``Quantum Anomalies and Chiral Magnetic Phenomena."
The work of D.O.R. and I.A.S. was supported by the U.S. National Science Foundation 
under Grants No. PHY-1404232 and No. PHY-1713950. The work of D.O.R. was also supported in part by
the 2018 Summer University Graduate Fellowship from the Department of Physics at Arizona State University. 
The work of E.V.G. was supported in part by the Program of Fundamental Research of the Physics and 
Astronomy Division of the National Academy of Sciences of Ukraine.
\end{acknowledgments}

\appendix

\section{Table integrals and thermodynamic functions in equilibrium}
\label{app:integrals}

In the calculation of moments of the distribution function, the following integrals are useful:
\begin{eqnarray}
\int \frac{d^4p}{(2\pi\hbar)^3} \delta(p^2) (p\cdot u)^n f_0 &=& - \frac{\Gamma(n+2)}{4\pi^2} T^{n+2} \sum_{\chi=\pm1}  \chi^{n+2} \textrm{Li}_{n+2}
 \left( -e^\frac{\chi\mu_\lambda}{T} \right) \equiv I_{n+2},\\
\int \frac{d^4p}{(2\pi\hbar)^3} \delta(p^2)  (p\cdot u)^n p^\alpha f_0 &=& u^\alpha I_{n+3},  \\
\int \frac{d^4p}{(2\pi\hbar)^3} \delta(p^2)  (p\cdot u)^n p^\alpha p^\beta f_0 &=& \left( -\frac{1}{3} g^{\alpha\beta} + \frac{4}{3} u^\alpha u^\beta \right) I_{n+4},  \\
\int \frac{d^4p}{(2\pi\hbar)^3} \delta(p^2)  (p\cdot u)^n p^\alpha p^\beta p^\gamma f_0 &=& \left( - g^{(\alpha\beta} u^{\gamma)} + 2 u^\alpha u^\beta u^\gamma \right) I_{n+5},\\
\int \frac{d^4p}{(2\pi\hbar)^3} \delta(p^2) (p\cdot u)^n p^\alpha p^\beta p^\gamma p^\delta f_0 &=& 
  \left( \frac{1}{5} g^{(\alpha\beta} g^{\gamma\delta)} - \frac{12}{5} g^{(\alpha\beta} u^{\gamma} u^{\delta)}
  + \frac{16}{5} u^\alpha u^\beta u^\gamma u^\delta \right) I_{n+6},
\end{eqnarray}
where $f_0$ is the equilibrium function at vanishing vorticity, i.e.,
\begin{equation}
		f_0 = \frac{1}{1 + e^{\textrm{sign}(p_0)(\varepsilon_{p,0}-\mu_\lambda)/T}}, 
\end{equation}
with $\varepsilon_{p,0} = p\cdot u$, 
and the round brackets in superscripts denote the symmetrization over all possible permutations 
of indices, e.g., $A^{(\alpha}B^{\beta}C^{\gamma)}\equiv  
       ( A^{\alpha}B^{\beta}C^{\gamma} 
      + A^{\alpha}B^{\gamma}C^{\beta} 
      + A^{\beta}B^{\alpha}C^{\gamma} 
      + A^{\beta}B^{\gamma}C^{\alpha} 
      + A^{\gamma}B^{\beta}C^{\alpha} 
      + A^{\gamma}B^{\alpha}C^{\beta}
)/3!$.

It is easy to check that the lower moments can be obtained from the higher ones multiplying the latter 
by the four-velocity $u^{\mu}$. As is easy to check, the explicit results for several lowest-order moments read
\begin{eqnarray}
 I_1 &=& \frac{\mu_\lambda}{4\pi^2\hbar^3},\label{eq:I1_calculated} \\
 I_2 &=& \frac{\mu_\lambda^2}{8\pi^2\hbar^3} + \frac{T^2}{24\hbar^3}, \label{eq:I2_calculated}\\
 I_3 &=& \frac{\mu_\lambda^3}{12\pi^2\hbar^3} + \frac{\mu_\lambda T^2}{12\hbar^3},  \label{eq:I3_calculated}\\
 I_4 &=& \frac{\mu_\lambda^4}{16\pi^2\hbar^3} + \frac{\mu_\lambda^2 T^2}{8\hbar^3} + \frac{7\pi^2 T^4}{240\hbar^3}.
\label{eq:I4_calculated}
\end{eqnarray}
Note that these moments satisfy the following recurrent relation: $\partial I_{n+1}/\partial\mu_\lambda=n I_n $. 
Using this relation it is easy to obtain similar chains of integrals with the first derivative of the distribution function 
$f^{\prime}_0=\partial f_0/\partial \varepsilon_{p,0} = - \partial f_0/\partial\mu_\lambda$ 
if one makes a substitution 
$I_n \rightarrow -(n-1)I_{n-1}$. For $f^{\prime\prime}_0$, the substitution is $I_n \rightarrow (n-1)(n-2)I_{n-2}$ 
and so on.

By making use of the table integrals above, we derive the following equilibrium expressions for the key 
thermodynamics functions:
\begin{eqnarray}
\label{eq:hydro-n-eq}
	n_{\textrm{eq}} &=& \frac{\mu(\mu^2 + 3\mu_5^2 + \pi^2 T^2)}{3\pi^2\hbar^3},
	\\
\label{eq:hydro-n5-eq}
	n_{5,\textrm{eq}} &=& \frac{\mu_5(\mu_5^2 + 3\mu^2 + \pi^2 T^2)}{3\pi^2\hbar^3},
	\\
\label{eq:hydro-e-eq}
	\epsilon_{\textrm{eq}} &=& \frac{\mu^4+6\mu^2\mu_5^2+\mu_5^4}{4\pi^2\hbar^3}+\frac{T^2(\mu^2+\mu_5^2)}{2\hbar^3}+\frac{7\pi^2T^4}{60\hbar^3}.
\end{eqnarray}
Note also that $P_{\textrm{eq}} = \epsilon_{\textrm{eq}}/3$. In addition, we also find that, in equilibrium, the following 
nondissipative contributions to the currents and momentum density are nonzero
\begin{eqnarray}
	\nu_{\textrm{eq}}^\mu &=& \sigma_\omega \omega^\mu + \sigma_B B^\mu ,
	\\
	\nu_{5,\textrm{eq}}^\mu &=& \sigma^5_\omega \omega^\mu + \sigma^5_B B^\mu ,
	\\
	h_{\textrm{eq}}^\mu &=& \xi_\omega \omega^\mu + \xi_B B^\mu.
\end{eqnarray}
As is easy to see, these include the celebrated chiral magnetic \cite{Fukushima:2008xe,Vilenkin:1980fu,Metlitski:2005pr} 
and vortical \cite{Vilenkin:1978hb,Vilenkin:1979ui,Vilenkin:1980zv,Erdmenger:2008rm,Banerjee:2008th,Son:2009tf,
Sadofyev:2010pr,Neiman:2010zi,Landsteiner:2011cp} effects, as well as their energy counterparts. 
The corresponding anomalous transport coefficients are given by
\begin{eqnarray}
\label{eq:anomalous_coeffs-1}
	\sigma_\omega = \frac{\mu\mu_5}{\pi^2\hbar^2},
	&\qquad&
	\sigma_B = e\frac{\mu_5}{2\pi^2\hbar^2},
	\\
\label{eq:anomalous_coeffs-2}
	\sigma^5_\omega = \frac{3(\mu^2+\mu_5^2)+\pi^2 T^2 }{6\pi^2\hbar^2},
	&\qquad&
	\sigma^5_B =  e\frac{\mu}{2\pi^2\hbar^2},
	\\
\label{eq:anomalous_coeffs-3}
	\xi_\omega = \frac{\mu_5\left(\mu_5^2+3\mu^2+\pi^2 T^2\right)}{3\pi^2\hbar^2},
	&\qquad&
	\xi_B = e\frac{\mu\mu_5}{2\pi^2\hbar^2}.
\end{eqnarray}

\section{Bessel functions}
\label{app:Bessel}

In this appendix, we present some useful relations for the Bessel functions that are needed
for the analysis in the main text of the paper. 

Let us remember that the Bessel functions $J_m(z)$ have an infinite number of positive 
real zeros at $z=\alpha_{m,i}$, where $i=1,2,...$. (Note that we use a nonstandard 
notation $\alpha_{m,i}$ instead of the usual $j_{m,i}$.) 

By making use of the table integrals \cite{Gradshtein-Ryzhik}
\begin{eqnarray}
	\int_0^1 dx\; rJ_\nu(ax)J_\nu(bx) &=& \frac{bJ_{\nu-1}(b)J_\nu(a) - aJ_{\nu-1}(a)J_\nu(b)}{a^2-b^2} = \frac{aJ_{\nu+1}(a)J_\nu(b) - bJ_{\nu+1}(b)J_\nu(a)}{a^2-b^2}, \\
	\int_0^1 dx\; xJ^2_\nu(ax) &=& \frac{1}{2}\left( J_\nu^2(a) - \frac{2\nu}{a}J_\nu(a)J_{\nu-1}(a) + J^2_{\nu-1}(a)\right) = \frac{1}{2}\left( J_\nu^2(a) - \frac{2\nu}{a}J_\nu(a)J_{\nu+1}(a) + J^2_{\nu+1}(a)\right),
\end{eqnarray}
one can easily derive the following orthogonality relation:
\begin{equation} 
\label{relation1}
	\int_0^1 dx\; xJ_{\tilde m}(\alpha_{m,i} x)J_{\tilde m}(\alpha_{m,j} x) = \delta_{ij} \frac{1}{2} J^2_{m\pm 1}(\alpha_{m,i}), 
\end{equation}
which is valid for any $\tilde m = m-1, m, m+1$. In this connection, it is useful to note that 
$J_{m-1}(\alpha_{m,i})=-J_{m+1}(\alpha_{m,i})$. As is easy to check, the latter follows from 
the well-known recurrence relation \cite{Gradshtein-Ryzhik},
\begin{equation} 
	xJ_{\nu-1}(x) + xJ_{\nu+1}(x) = 2\nu J_\nu(x). 
\end{equation}
After integrating Eq.~(\ref{relation1}) by parts and taking into account the property of the 
Bessel functions \cite{Gradshtein-Ryzhik},
\begin{equation} 
	J_{\nu-1}(x) - J_{\nu+1}(x) = 2 J'_\nu(x).
\end{equation}
we easily derive the following two integral relations:
\begin{eqnarray} 
\label{relation2}
	\left(\frac{1}{2} J^2_{m\pm 1}(\alpha_{m,i})\right)^{-1} \int_0^1 dx\; x^2J_m(\alpha_{m,i} x)J_{m\pm 1}(\alpha_{m,i} x) 
	&=& \frac{m\pm 1}{\alpha_{m,i}}, \\
\label{relation3}
	\left(\frac{1}{2} J^2_{m\pm 1}(\alpha_{m,i})\right)^{-1} \int_0^1 dx\; x^3J_{\tilde m}(\alpha_{m,i} x)J_{\tilde m}(\alpha_{m,i} x) 
	&=& \frac{1}{3\alpha_{m,i}^2}\left( \frac{2\tilde m(\tilde m^2 - 1)}{m} + \alpha_{m,i}^2 \right),
\end{eqnarray}
where again $\tilde m = m-1, m, m+1$.

\section{Explicit form of linearized equations}
\label{app:Full_system}

In this appendix, we present the explicit form of the linearized equations for the perturbations around the 
unperturbed equilibrium state of a uniformly rotating charged chiral plasma. 

By making use of Eqs.~(\ref{eq:continuity-1})--(\ref{eq:maxwell}) and the ansatz in 
Eqs.~(\ref{eq:perturbations-1})--(\ref{eq:perturbations-3}) for the perturbations of the plasma 
parameters, we can derive the coupled system of linearized equations. The electric charge 
conservation relation leads to 
\begin{eqnarray} 
	&\sum_i \left[ 
	-i(k_0-m\Omega)\frac{\partial n}{\partial \zeta_i} 
	- i k_\mu \omega^\mu \frac{\partial \sigma_\omega}{\partial \zeta_i} 
	- i k_\mu \left(B^\mu - \frac{1}{2}en\omega^\mu r^2\right) \frac{\partial \sigma_B}{\partial \zeta_i}
	+ \frac{\tau}{3}\frac{\partial n}{\partial \zeta_i} \left( (k_0-m\Omega)^2 - k_\mu k^\mu - \partial_r^2 - \frac{1}{r}\partial_r + \frac{m^2}{r^2} \right)
	\right]\delta \zeta_i
	\nonumber
	\\
	&\qquad
	+ n\left[ 1+i\tau(k_0-m\Omega) \right] \left( -ik_\mu\delta u^\mu + D_1[\delta u] \right)
	- 2i\Omega \tau n D_2[\delta u]
	+ \frac{1}{e} \sigma_E \left( -ik_\mu\delta E^\mu + D_1[\delta E] \right) = 0 .
\end{eqnarray}
Similarly, the chiral charge conservation is given by
\begin{eqnarray}
	&\sum_i \left[ 
	-i(k_0-m\Omega)\frac{\partial n_5}{\partial \zeta_i} 
	- i k_\mu \omega^\mu \frac{\partial \sigma^5_\omega}{\partial \zeta_i} 
	- i k_\mu \left(B^\mu - \frac{1}{2}en\omega^\mu r^2\right) \frac{\partial \sigma^5_B}{\partial \zeta_i}
	+ \frac{\tau}{3}\frac{\partial n_5}{\partial \zeta_i} \left( (k_0-m\Omega)^2 - k_\mu k^\mu - \partial_r^2 - \frac{1}{r}\partial_r + \frac{m^2}{r^2} \right)
	\right]\delta \zeta_i
	\nonumber
	\\
	&\qquad
	+ 2i \sigma^5_\omega k_0\omega^\mu \delta u_\mu
	+ \sigma^5_B \left( -ik_\mu\delta B^\mu + D_1[\delta B] \right) = - \frac{e^2}{2\pi^2\hbar^2} \delta E^\mu \left(B_\mu - \frac{1}{2}en\omega_\mu r^2\right),
\end{eqnarray}
The energy-momentum conservation relations read
\begin{eqnarray}
	&\sum_i \left[ 
	-i(k_0-\frac{4}{3}m\Omega)\frac{\partial \epsilon}{\partial \zeta_i}
	- ik^\mu\omega_\mu \frac{\partial \xi_\omega}{\partial \zeta_i}
	- ik^\mu\left(B_\mu - \frac{1}{2}en\omega_\mu r^2\right) \frac{\partial \xi_B}{\partial \zeta_i}
	\right] \delta \zeta_i
	\nonumber
	\\
	&\qquad
	+ \frac{4}{3}\epsilon \left[ -i k^\mu\delta u_\mu + D_1[\delta u] - k_0\Omega r \left(\delta u^+ - \delta u^-\right) \right]
	- \frac{4\tau\epsilon}{45}im\Omega \left(i k_z\delta u^3 + D_1[\delta u] \right)
	\nonumber
	\\
	&\qquad
	- \frac{4\tau\epsilon}{15} i\Omega r \left[ k_z^2 - \partial_r^2 - \frac{1}{r}\partial_r \right]\left(\delta u^+ - \delta u^-\right)
	- \frac{4\tau\epsilon}{15} i\Omega r \left[\frac{(m+1)^2}{r^2}\delta u^+ - \frac{(m-1)^2}{r^2} \delta u^-\right]
	\nonumber
	\\
	&= \sum_i \left[ 
	e \frac{\tau}{3}\frac{\partial n}{\partial \zeta_i} B\Omega r \partial_r
	\right] \delta \zeta_i
	- ik_0 e\tau n B\Omega r \left(\delta u^+ + \delta u^-\right)
	- \sigma_E B\Omega r \left(\delta E^+ + \delta E^-\right)
	- ien\Omega r \left(\delta E^+ - \delta E^-\right),
\end{eqnarray}
\begin{eqnarray} 
	&\sum_i \left[ 
	\frac{1}{6}\frac{\partial \epsilon}{\partial \zeta_i}\left(\partial_r - s\frac{m}{r} + 4sk_0 \Omega r\right)
	- \frac{1}{2} sk_z B\Omega r\frac{\partial \xi_B}{\partial \zeta_i}
	\right] \delta \zeta_i
	+ \frac{2}{3}is\epsilon\Omega r \left( -i k^\mu\delta u_\mu + D_1[\delta u] \right)
	\nonumber
	\\
	&\qquad
	- \frac{4}{3}i\epsilon \left[k_0 - (m+2s)\Omega\right]\delta u^s
	- \frac{2\tau\epsilon}{45} \left[ \partial_r - s\frac{m}{r} + sk_0\Omega r\right]\left( ik_z\delta u^3 + D_1[\delta u]\right)
	\nonumber
	\\
	&\qquad
	+ \frac{2\tau\epsilon}{45} k_0\Omega r\left[ \partial_r - s\frac{m-s}{r} \right]\left(\delta u^+ - \delta u^-\right)
	+ \frac{4\tau\epsilon}{15} \left[ k_z^2 - \partial_r^2 - \frac{1}{r}\partial_r + \frac{(m+s)^2}{r^2} \right] \delta u^s
	- \frac{8\tau\epsilon}{15} k_0(m+2s)\Omega\delta u^s
	\nonumber
	\\
	&= \sum_i \left[ 
	e \frac{\tau}{6}is\left(B - \frac{1}{2}en\Omega r^2\right)\frac{\partial n}{\partial \zeta_i} \left(\partial_r - s\frac{m}{r} \right)
	+ e\frac{\tau}{6} ik_0 B\Omega r \frac{\partial n}{\partial \zeta_i}
	\right] \delta \zeta_i
	+ es \left(B - \frac{1}{2}en\Omega r^2\right) \left[\tau k_0 n\delta u^s 
	- \frac{i}{e}\sigma_E \delta E^s \right]
	+ en \delta E^s,
\end{eqnarray}
\begin{eqnarray} 
	&\sum_i \left[ 
	\frac{1}{3}ik_z\frac{\partial \epsilon}{\partial \zeta_i}
	- ik_0\Omega \frac{\partial \xi_\omega}{\partial \zeta_i}
	- ik_0\left(B - \frac{1}{2}en\Omega r^2\right) \frac{\partial \xi_B}{\partial \zeta_i}
	+ im\Omega B \frac{\partial \xi_B}{\partial \zeta_i}
	\right] \delta \zeta_i
	-\frac{4}{3}i\epsilon (k_0-m\Omega) \delta u^3
	- \frac{8\tau\epsilon}{15} k_0 m\Omega\delta u^3
	\nonumber
	\\
	&\qquad
	+ \frac{4\tau\epsilon}{45} \left( k_z^2\delta u^3 - ik_z D_1[\delta u] + ik_zk_0\Omega r \left(\delta u^+ - \delta u^-\right)\right)
	+ \frac{4\tau\epsilon}{15} \left[ k_z^2 - \partial_r^2 - \frac{1}{r}\partial_r + \frac{m^2}{r^2} \right] \delta u^3
	= en\delta E^3.
\end{eqnarray}
Finally, the Maxwell equations take the form 
\begin{eqnarray} 
	- 2\Omega\delta E^3 
	+ ik\Omega r \left(\delta E^+ + \delta E^-\right)
	- \Omega r \partial_r \delta E^3
	- i(k_0 - m\Omega) \delta B^0
	+ i k^\mu\delta B_\mu - D_1[\delta B]
	- ik_zB\delta u^0 &=& 0,
	\\
	- k_z s \delta E^s
	- \frac{1}{2}is \left[ \partial_r - s\frac{m}{r} \right]\delta E^3
	- \frac{1}{2}ik_0\Omega r\delta E^3
	- i\left(k_0 - m\Omega\right) \delta B^s 
	\nonumber\\
	- \frac{1}{2}is\Omega r\left( - i k^\mu\delta B_\mu + D_1[\delta B] \right)
	- ik_z \left(B - \frac{1}{2}en\Omega r^2\right)\delta B^s
	&=& 0,
	\\
	- i D_2[\delta E]
	+ ik_0\Omega r \left(\delta E^+ + \delta E^-\right)
	+ \left(B - \frac{1}{2}en\Omega r^2\right)\left( -ik_0 \delta u^0 + D_1[\delta u] \right)
	\nonumber\\
	- i(k_0 - m\Omega)\delta B^3
	- en\Omega r \left(\delta u^+ + \delta u^-\right)
	&=& 0,
\end{eqnarray}
and
\begin{eqnarray} 
	&- i\left(B - \frac{1}{2}en\Omega r^2\right)D_2[\delta u]
	+ 2\Omega \delta B^3
	+ ien \Omega r \left(\delta u^+ - \delta u^-\right)
	- ik_z \Omega r \left(\delta B^+ + \delta B^-\right)
	+ \Omega r\partial_r\delta B^3
	\nonumber
	\\
	&\qquad
	- ik_3 \delta E^3 - D_1[\delta E]
	=
	e\sum_i \left[ 
	\left( 1 - i\frac{\tau}{3}\Omega m \right)\frac{\partial n}{\partial \zeta_i}
	\right] \delta \zeta_i
	+ en(1 + i\tau k_0)\delta u^0
	+ \sigma_E \delta E^0,
\end{eqnarray}
\begin{eqnarray} 
	&sk_0\left(B - \frac{1}{2}en\Omega r^2\right)\delta u^s
	+ \frac{1}{2}sB\Omega r\left[ \partial_r - s\frac{m-s}{r} \right]\left(\delta u^+ - \delta u^-\right)
	- \frac{1}{2}iBk_z\Omega r \delta u^3
	+ sk_z \delta B^s
	\nonumber
	\\
	&\qquad
	+ \frac{1}{2}is \left[ \partial_r - s\frac{m}{r} \right] \delta B^3
	+ \frac{1}{2}ik_0\Omega r \delta B^3
	- i\left( k_0-m\Omega \right) \delta E^s
	- is\Omega r \left( -ik_\mu\delta E^\mu + D_1[\delta E] \right)
	\nonumber
	\\
	&\qquad
	= en\delta u^s
	+ ie\tau n\left[k_0 - (m+2s)\Omega \right] \delta u^s
	+ e\sum_i \left[ 
	\frac{1}{2}is\Omega r\frac{\partial n}{\partial \zeta_i}
	- \frac{\tau}{6}\frac{\partial n}{\partial \zeta_i} \left( \partial_r - s\frac{m}{r} + sk_0\Omega r \right)
	\right] \delta \zeta_i
	+ \sigma_E \delta E^s,
\end{eqnarray}
\begin{eqnarray} 
	&iD_2[\delta B]
	- ik_0\Omega r \left(\delta B^+ + \delta B^-\right)
	+ B\Omega r\left[ \partial_r + \frac{2}{r} \right]\delta u^3
	- i(k_0 -m\Omega)\delta E^3
	\nonumber
	\\
	&\qquad
	= e\sum_i \left[ 
	- \frac{\tau}{3}ik_z\frac{\partial n}{\partial \zeta_i}
	+ \Omega \frac{\partial \sigma_\omega}{\partial \zeta_i}
	+ \left(B - \frac{1}{2}en\Omega r^2\right)\frac{\delta\sigma_B}{\partial \zeta_i}
	\right] \delta \zeta_i
	+ en\delta u^3
	+ ie(k_0-m\Omega)\tau n\delta u^3
	+ \sigma_E \delta E^3 .
\end{eqnarray}
In these equations, the index $s=\pm 1$ labels circular polarizations and the sums over $i=1,2,3$ account 
for the variations of the three physical parameters, $\delta \zeta_i = \delta\mu, \delta\mu_5, \delta T$. 
Note that the variations of all quantities are assumed to have a radial dependence, i.e.,
$\delta\mu = \delta\mu(r), \delta T=\delta T(r)$, etc., although it is not shown explicitly. In the 
linearized equations, we used the following differential operators:
\begin{eqnarray} 
	D_1[\delta v] &=& \partial_r (\delta v^+ + \delta v^-) + \frac{m+1}{r}\delta v^+ - \frac{m-1}{r}\delta v^-,
	\\
	D_2[\delta v] &=& \partial_r (\delta v^+ - \delta v^-) + \frac{m+1}{r}\delta v^+ + \frac{m-1}{r}\delta v^-.
\end{eqnarray}
Last but not least, let us remind the reader that the zeroth components of the vector quantities 
are not independent. They are expressed in terms of the spatial component; see 
Eqs.~(\ref{eq:perturb-u0})--(\ref{eq:perturb-B0}).

\section{Damping of the chiral magnetic wave}
\label{noCMW}

In this appendix, we present a detailed explanation of why there is no propagating chiral magnetic 
wave when dynamical electromagnetism is properly accounted for in the analysis. The underlying 
argument, as we will see, is rather general and very simple at the same time. From a physics 
viewpoint, the chiral magnetic wave becomes overdamped because of a high conductivity of 
chiral plasma, which causes a rapid screening of electric charge fluctuations. Therefore, 
the would-be chiral magnetic wave cannot form and cannot sustain itself.

For the purposes of understanding how dynamical electromagnetism turns the chiral magnetic 
wave into an overdamped mode, it is sufficient to consider the system without rotation, 
$\Omega =0$. For the sake of simplicity, here we will concentrate on the case
of hot plasma with the vanishing average $\mu$ and $\mu_5$ in equilibrium.
The corresponding (linearized) hydrodynamic equations are given in 
Eqs.~(\ref{eq:system-simple-start})--(\ref{eq:system-simple-end-hydro}). They 
are also supplemented by the Maxwell equations (\ref{Maxwell111}) and (\ref{eq:system-simple-end}).
In general, the complete system of equations does not decouple into separate subsystems and the 
analysis is rather complicated. One of the rare exceptions happens to be the system of 
equations that describes the longitudinal chiral magnetic wave with $\mathbf{k}\parallel \mathbf{B}_0 $, 
where $\mathbf{k}$ is the wave vector and $\mathbf{B}_0 $ is the background magnetic field.

Let us start by reminding the reader that the existence of the chiral magnetic wave \cite{Kharzeev:2010gd}
stems from the interplay between the fermion number (or electric charge) and chiral charge fluctuations.
Such fluctuations must satisfy the continuity relations (\ref{eq:system-simple-start}) and (\ref{eq:system-simple-start001}). In 
the rest frame of plasma, the corresponding equations take the following explicit form: 
\begin{eqnarray}
\label{eq1}
\frac{\partial \delta n}{\partial t} - \frac{\tau}{3}\bm{\nabla}^2\delta n
+ \bm{\nabla}\cdot \left( \frac{e}{2\pi^2 \hbar^2 \chi_5}\mathbf{B}_0  \delta n_5 
+\frac{1}{e}\sigma_E \delta\mathbf{E}\right) &=& 0,\\
\label{eq2}
\frac{\partial \delta n_5}{\partial t} -\frac{\tau}{3}\bm{\nabla}^2\delta n_5 
+ \bm{\nabla}\cdot \left( \frac{e}{2\pi^2 \hbar^2 \chi }\mathbf{B}_0  \delta n \right) -
\frac{e^2}{2\pi^2\hbar^2} (\mathbf{B}_0 \cdot \delta \mathbf{E}) &=& 0 ,
\end{eqnarray}
where we introduced two types of susceptibilities, i.e., $\chi \equiv \delta n / \delta \mu$ and 
$\chi_5 \equiv \delta n_5 / \delta\mu_5$, which can be obtained from the explicit expressions 
for the densities in Eqs.~(\ref{eq:hydro-n-eq}) and (\ref{eq:hydro-n5-eq}). It should 
be emphasized that these equations do not contain any dependence on the oscillations of the 
fluid velocity $\delta \mathbf{u}$. While this is not a general property, it can be shown to be true 
in the special case of a longitudinal wave (i.e., $\mathbf{k}\parallel \mathbf{B}_0 $), which we assumed
in writing down Eqs.~(\ref{eq1}) and (\ref{eq2}). 

In the analysis here, we include Ohm's current determined by the electrical conductivity $\sigma_E$ in Eq.~(\ref{eq1}), 
as well as the anomalous contribution $\propto (\mathbf{B}_0 \cdot \delta \mathbf{E})$ in Eq.~(\ref{eq2}). 
In a naive background field approximation, which is often used in the literature, both of these 
effects are usually ignored. As we will demonstrate below, however, they are not negligible and, in fact, 
play important roles in the dynamics of the collective mode in question. 

After using Gauss's law, $\bm{\nabla}\cdot \delta\mathbf{E} = e \delta n$, the above set of equations 
can be rewritten as follows:
\begin{eqnarray}
\label{eq10}
\frac{\partial \delta n}{\partial t} - \frac{\tau}{3}\bm{\nabla}^2\delta n
+ \frac{e}{2\pi^2 \hbar^2 \chi_5}\left( \mathbf{B}_0  \cdot \bm{\nabla} \right)   \delta n_5+\sigma_E \delta n&=& 0,\\
\label{eq20}
\frac{\partial \delta n_5}{\partial t} -\frac{\tau}{3}\bm{\nabla}^2\delta n_5 
+  \frac{e}{2\pi^2 \hbar^2 \chi }\left( \mathbf{B}_0  \cdot \bm{\nabla} \right)   \delta n
- \frac{e^2}{2\pi^2\hbar^2} (\mathbf{B}_0 \cdot \delta \mathbf{E}) &=& 0 .
\end{eqnarray}
Furthermore, by assuming that the oscillations of densities and the electric field are in the form of plane waves
$\propto e^{-i k_0 t +i \mathbf{k}\cdot \mathbf{r}}$, we derive
\begin{eqnarray}
\label{eq11}
k_0 \delta n  + i \frac{\tau}{3} k^2\delta n
 - \frac{eB_0 }{2\pi^2 \hbar^2 \chi_5}k \delta n_5  + i \sigma_E \delta n &=& 0 ,\\
\label{eq22}
k_0 \delta n_5  + i \frac{\tau}{3}k^2\delta n_5 
- \frac{eB_0 }{2\pi^2 \hbar^2 \chi }k \delta n - 
\frac{e^3B_0}{2\pi^2\hbar^2} \frac{1}{k}  \delta n&=& 0 .
\end{eqnarray}
where, in the last equation, we took into account that $(\mathbf{B}_0 \cdot \delta \mathbf{E}) = 
B_0(\mathbf{k}\cdot \delta \mathbf{E})/k  = -i (e B_0/k) \delta n $ for a longitudinal wave. In order to see that 
Eqs.~(\ref{eq11}) and (\ref{eq22}) do not give any propagating modes, it is sufficient to use simple 
parametric estimates for the key terms in the equations. In the regime of hot plasma, for example, 
we simply need to know that $\chi\approx \chi_5\sim T^2$ and $\sigma_E \sim e^2 T^2  \tau \sim T/(e^2 \ln e^{-1})$. 
In the last relation, we estimated the transport time as $\tau \sim 1/(e^4 T \ln e^{-1})$ \cite{Baym:1990uj,
Arnold:2000dr}. (While we concentrate 
primarily on the regime of hot plasma in this appendix, similar arguments can be given also in
the case of dense plasma. The corresponding estimates will be similar, but the role of the hard energy
scale will be played by the chemical potential rather than temperature, i.e., $\chi \approx \chi_5\sim \mu^2$ 
and $\sigma_E \sim \mu/e^2$. As for the key qualitative conclusions, they will remain the same.)
 
If the electrical conductivity term in Eq.~(\ref{eq11}) were neglected (i.e., if we set $\sigma_E=0$ by hand), 
the longitudinal chiral magnetic wave would proceed in the usual way: an oscillation of $\delta n_5$ would
induce a nonzero $\delta n$ via the chiral magnetic effect and that, in turn, would drive again a nonzero 
oscillation of $\delta n_5$ via the chiral separation effect in Eq.~(\ref{eq22}). It is interesting to note that 
dynamical electromagnetism still plays an important role even when the conductivity effects are neglected. 
Indeed, it gives rise to the anomalous term $\propto (e^3B_0/k)$ in the second equation, which is responsible
for a nonzero gap in the dispersion relation,
\begin{eqnarray}
\label{mod01}
k_0^{(\pm)} &=& \pm \frac{e^2 B_0 }{2\pi^2\hbar^2\sqrt{\chi_5}} \sqrt{1+\frac{k^2}{e^2\chi}}  -i \frac{\tau}{3}k^2 , \quad
\mbox{for}\quad \sigma_E=0.
\end{eqnarray}
It should be clear that the diffusion effects $\propto  -i \tau k^2/3$ are small for the modes with sufficiently 
large wavelengths. Therefore, this is a well resolved propagating mode with a nonzero gap $\propto e^2 B_0/T$. 
As is easy to check, the gap would vanish if the last anomalous term in Eq.~(\ref{eq2}) were absent. 
Note that the same gapped mode was previously found in the context of Weyl semimetals in 
Ref.~\cite{Gorbar:2018nmg}.

The propagation of the chiral magnetic wave is badly disrupted, however, when the effects of electrical 
conductivity are properly accounted for. In essence, because of a very rapid screening of local fluctuations 
of the electric charge, i.e., $\delta n \propto e^{-\sigma_E t}$, the wave does not get a chance to form. 
This is also clear from Eq.~(\ref{eq11}) where the term with electrical conductivity is much larger than 
the anomalous term $\propto (eB_0 /\chi_5) k \delta n_5$. Indeed, for all realistic values of the magnetic 
field and wave vectors, $eB_0  k \ll \sigma_E \chi_5  \sim T^3/e^2$. The dispersion relations for the resulting 
dissipative modes are given by the following explicit expressions: 
\begin{eqnarray}
\label{mod01}
k_0^{(\pm)} &=& -\frac{i}{2}\sigma_E 
\pm \frac{i}{2}\sigma_E \sqrt{1-\frac{e^4 B_0^2}{\pi^4\hbar^4\sigma_E^2\chi_5}
\left(1+\frac{k^2}{e^2\chi}\right)}  -i \frac{\tau}{3}k^2 ,
\end{eqnarray}
This result confirms that the effects of high electrical conductivity are responsible for turning the chiral magnetic 
waves into overdamped modes. From a physics viewpoint, it is the screening of charge fluctuations in the 
plasma that prevents the wave from propagating. 

In conclusion, the arguments presented here are quite general and should remain valid in almost any regime 
of a relativistic plasma. The only regime where this may not be true is, perhaps, the ultraquantum limit with 
$|eB_0 |\gg T^2$. In such an exceptional case, however, the lowest Landau level dominates the dynamics 
and the usual hydrodynamic description becomes inapplicable.

\end{document}